\documentclass[sigconf]{acmart} % two column
\usepackage{graphicx} % ``demo`` option just for this example
\usepackage{subcaption}

\usepackage{multirow}
\usepackage{xcolor,colortbl}
\usepackage{balance}
\usepackage{flushend}
\usepackage{soul}
\usepackage{changepage}

\definecolor{OliveGreen}{rgb}{0,0.6,0}
\definecolor{lightred}{rgb}{1, 0.8, 0.8}
\definecolor{lightgreen}{rgb}{0.8, 1, 0.8}
\definecolor{blue}{rgb}{0, 0, 0}

\usepackage[suppress]{color-edits} % Uncomment for final draft
\addauthor{ls}{olive} % Logan
\addauthor{hz}{purple} % Haiyi
\addauthor{ac}{brown} % Alex
\addauthor{kh}{red} % Ken
\addauthor{sw}{blue} % Steven
\addauthor{ml}{orange} % Min
\addauthor{dq}{green} % Diana
\addauthor{mw}{pink} % Marya

\newcommand{\xhdr}[1]{\vspace{2mm} \noindent{\bf #1}}

\usepackage{tikzsymbols}

\usepackage{etoolbox}
\AtBeginDocument{%
  \providecommand\BibTeX{{%
    \normalfont B\kern-0.5em{\scshape i\kern-0.25em b}\kern-0.8em\TeX}}}
    
\keywords{child welfare; machine learning; participatory design; human-centered AI; impacted stakeholder}

\copyrightyear{2022}
\acmYear{2022}
\setcopyright{acmlicensed}\acmConference[FAccT '22]{2022 ACM Conference on Fairness, Accountability, and Transparency}{June 21--24, 2022}{Seoul, Republic of Korea}
\acmBooktitle{2022 ACM Conference on Fairness, Accountability, and Transparency (FAccT '22), June 21--24, 2022, Seoul, Republic of Korea}
\acmPrice{15.00}
\acmDOI{10.1145/3531146.3533177}
\acmISBN{978-1-4503-9352-2/22/06}

\begin{document}

%%
%% The ``title`` command has an optional parameter,
%% allowing the author to define a ``short title`` to be used in page headers.

\title[Imagining new futures beyond predictive systems in child welfare]{Imagining new futures beyond predictive systems in child welfare: A qualitative study with impacted stakeholders}

\author{Logan Stapleton}
\affiliation{%
  \institution{University of Minnesota}
  \city{Minneapolis}
  \country{USA}
  \orcid{0000-0003-3373-1048}
}
\email{stapl158@umn.edu}

\author{Min Hun Lee}
\authornote{Lee completed much of this work while at Carnegie Mellon University.}
\affiliation{%
  \institution{Singapore Management University}
  \country{Singapore}
}

\author{Diana Qing}
\affiliation{%
  \institution{University of California, Berkeley}
  \city{Berkeley}
  \country{USA}
}

\author{Marya Wright}
\affiliation{%
  \institution{University of Southern California}
  \city{Oakland}
  \country{USA}
}

\author{Alexandra Chouldechova}
\affiliation{%
  \institution{Carnegie Mellon University}
  \city{Pittsburgh}
  \country{USA}
}

\author{Kenneth Holstein}
\affiliation{%
  \institution{Carnegie Mellon University}
  \city{Pittsburgh}
  \country{USA}
}

\author{Zhiwei Steven Wu}
\affiliation{%
  \institution{Carnegie Mellon University}
  \city{Pittsburgh}
  \country{USA}
}

\author{Haiyi Zhu}
\affiliation{%
  \institution{Carnegie Mellon University}
  \city{Pittsburgh}
  \country{USA}
}

\renewcommand{\shortauthors}{Stapleton et al.}

\begin{abstract}
Child welfare agencies across the United States are turning to data-driven predictive technologies (commonly called predictive analytics) which use government administrative data to assist workers' decision-making. While some prior work has explored impacted stakeholders’ concerns with current uses of data-driven predictive risk models (PRMs), less work has asked stakeholders whether such tools ought to be used in the first place. In this work, we conducted a set of seven design workshops with 35 stakeholders who have been impacted by the child welfare system or who work in it to understand their beliefs and concerns around PRMs, and to engage them in imagining new uses of data and technologies in the child welfare system. We found that participants worried current PRMs perpetuate or exacerbate existing problems in child welfare. Participants suggested new ways to use data and data-driven tools to better support impacted communities and suggested paths to mitigate possible harms of these tools. Participants also suggested low-tech or no-tech alternatives to PRMs to address problems in child welfare. Our study sheds light on how researchers and designers can work in solidarity with impacted communities, possibly to circumvent or oppose child welfare agencies.
\end{abstract}

\maketitle

\section{Introduction}
\label{sec:intro}
\textit{Where should we send the police? Who should we give housing to? How should we educate our children? Who should we give unemployment benefits to? Which families should we investigate for child abuse?} AI-based predictive algorithms are being used or are being considered for use across all of these everyday public sector decisions \cite{holstein2018student,chouldechova2018case,brayne2017policing,toros2018homeless,panoptykon2015unemployed}. Many of these technologies have faced public scrutiny and opposition. For example, in St. Paul, Minnesota, an algorithm intended to assess which children were at risk of getting involved in the juvenile justice system was blocked by a group of impacted parents and teachers who organized to oppose it \cite{pomeroy2019community}. While some government agencies have established track records of community engagement around the deployment of new technologies, the perspectives of stakeholders who will be most impacted by algorithms are not always adequately considered \cite{brown2019toward,holtenmoller2020shifting,robertson2020if,zhu2018value}.

In this paper, we aim to address the following research question: \textit{What do impacted stakeholders think about data-driven technologies in the child welfare system?} To do so, we held seven workshops with 35 expert stakeholders who are personally impacted by child protective services (CPS) and/or work in CPS. We first explained to our participants how current data-driven predictive risk models (henceforth \textit{PRMs}) are designed and used. We then talked with participants about their perspectives on these technologies. We also encouraged participants to weigh in on whether current PRMs address the main problems they see in CPS, and to imagine other possibilities for data and data-driven tools beyond current PRMs. Prior work with impacted stakeholders has explored the design and use of PRMs \cite{brown2019toward}. Our study is the first in academic ML and HCI to ask stakeholders whether these technologies should be used at all and to imagine new futures beyond them. Yet, these conversations have been ongoing outside these academic disciplines \cite{webeimagining2020mothers,endup2021}.\footnote{See, e.g., the 2021 upEND Movement Convening keynote with Derecka Purnell and Dorothy Roberts: \url{https://youtu.be/udIq9oRDcDQ}.}

Our participants brought up several important themes: In Section~\ref{sec:participant-concerns}, we note that most participants opposed current PRMs because they saw them as exacerbating existing problems in CPS. These findings are consistent with, yet more specific and more critical than, prior work \cite{brown2019toward}. We present these first as a primer to more novel, constructive suggestions in Sections~\ref{sec:new-uses}, ~\ref{sec:guidelines}, and \ref{sec:low-tech-alternatives}. In Section~\ref{sec:new-uses}, we present participants' suggestions for new data-driven tools beyond PRMs which better support impacted communities, e.g. to evaluate the child welfare system and the people who work in it, to recommend mandated reporters when \textit{not} to make a report, and to allocate resources to families to prevent child maltreatment. In Section~\ref{sec:guidelines}, participants recommended guidelines to mitigate possible harms of PRMs if they must be used in the future. In Section~\ref{sec:low-tech-alternatives}, participants suggested low-tech and no-tech alternatives better address the problems that motivate the use of PRMs. Overall, our work advances ongoing discussions around data-driven tools in CPS. We argue against current PRMs, and give new avenues to work in solidarity with impacted communities, beyond just designing algorithms for CPS agencies.

\section{Related Work}
\label{sec:related-work}

\subsection{Algorithms in child welfare}
CPS agencies have been using checklist-style actuarial risk assessments (henceforth \textit{diagnostic checklists}), such as Structured decision-making (SDM) \cite{sdm}, for decades to assess how likely they think a family is to harm their children. Many agencies also use \textit{practice models} such as Signs of Safety (SofS) and Safety Organized Practice (SOP) \cite{turnell1997aspiring} as decision-making guides, often in conjunction with diagnostic checklists \cite{mickelson2017assessing}. For a case study of diagnostic checklists, see \cite{saxena2021framework}. \citet{saxena2020human} note that predictive risk models (PRMs) which apply machine learning to administrative data have grown in popularity since around 2015. Some PRMs have been developed by private companies \cite{eckerdrsf,mindshare,sas}. However, due to high error rates and proprietary opacity, many have been dropped \cite{nccpr2017losangeles,nash2017losangeles,jackson2017illinois}. Other PRMs have been developed through public-academic partnerships \cite{vaithianathan2017,chouldechova2018case,riley2018can,douglascounty}. PRMs are currently being used or deployed in (at least) Pennsylvania, New York, Florida, Washington, Oregon, Colorado, and California \cite{aclu2021family}. For an extensive list of algorithms used in the U.S. child welfare system, see \cite{aclu2021family} or \cite{saxena2020human}. PRMs have been deployed in response to racial biases and disparities \cite{dettlaff2011disentangling,Kim2017lifetime}, inaccurate and inconsistent decisions, child fatalities \cite{netflix2020gabrielfernandez}, etc. Proponents of PRMs argue they make more accurate decisions than both workers and diagnostic checklists; and that they make more consistent, objective, and equitable decisions \cite{dare2016ethical,chouldechova2018case,hurley2018algorithm,stack2018cyf,dhs2019impactsummary}. Some critics disagree with these points, arguing that PRMs are still discriminatory and still too inaccurate \cite{eubanks2018automating,nccpr2018predictive,church2017silver}. Others argue that PRMs risk ``coding over the cracks'' without addressing the foundational flaws in child welfare, and that communities should instead organize around systemic improvements to address these flaws \cite{glaberson2019coding}. Others still argue that CPS is not a flawed system but a carceral one that plays a dual, paradoxical role \cite{pelton1994,roberts2002shattered,roberts2007paradox, copeland2021only} to police families while supporting them --- and that the supportive, ``welfare'' side is an over-stated veneer to cover up the real carceral side \cite{roberts2022torn}. These critics argue that PRMs introduce new ways for CPS to police Black, Indigenous, and poor families \cite{abdurahman2021calculating,roberts2019digitizing,roberts2022torn}.

\subsection{Participatory algorithm design}
Influenced by action research and the work of Paulo Freire \cite{freire1972pedagogy}, \textit{participatory design} developed around the 1970s by Scandinavian researchers working to gain workers more power over the design of technologies they use on the job \cite{kyng1979systems,sandberg1979computers,bjerknes1987computers,gregory2003scandinavian}. Participatory methods have since become a mainstay in HCI and CSCW \cite{muller1993participatory,kensing1998participatory}, but have been broadened beyond their Marxist roots \cite{spinuzzi2002scandinavian,bjorgvinsson2010participatory}. More recently, many have called for increased participation to ensure that diverse stakeholders' perspectives, needs, and values are reflected in the design of AI systems \cite{paml,loi2018pd,varshney2021participatory,pair2020boundary,zhu2018value,wong2020democratizing}. Yet, without clear political motivations beyond ``democratization'' of AI governance, participatory work in ML differs widely based on ``which stakeholders are involved'' and ``what is on the table'' \cite{delgado2021stakeholder,wolf2018changing,sloane2020participation}. Some propose consulting ``the public'' or broadly-defined ``stakeholders'' on their preferences around specific, technical design decisions \cite{lee2019webuildai,awad2018moral,noothigattu2018voting,kahng2019statistical,grgichlaca2018,ilvento2019metric,bechavod2020metric,jung2021algorithmic,johnston2020preference,robertson2020if}. Others intentionally work with specific groups who are most impacted by these technologies, yet still do not empower impacted stakeholders to engage in broader design decisions \cite{brown2019toward,holtenmoller2020shifting,saxena2020participatory,cheng2021soliciting,smith2020community,aragon2022human,smith2020community,halfaker2020ores}. While more common across HCI and CSCW, less work in participatory ML empowers stakeholders to decide on the ``scope and purpose for AI, including whether it should be built or not'' \cite{delgado2021stakeholder}. Specifically around the design of algorithms in child welfare,\footnote{This point might be broadened to public algorithms in general, e.g. \cite{holtenmoller2020shifting}. Though, forthcoming work centers people seeking government services \cite{scott2022algorithmic}.}  prior participatory work has either collaborated with government agencies or solely engaged with government workers in their studies \cite{brown2019toward,saxena2020participatory,kawakami2022partnerships,cheng2022disparities,kawakami2022exploring}.\footnote{Harding argues ``value-neutral'' sciences side with the powerful, e.g. ``the welfare department instead of the people who were receiving welfare'' \cite{harding2016standpoint}.} Most similar to our work, \citet{brown2019toward} partnered with a CPS agency to aid the development of a PRM by conducting participatory design workshops where they asked workers and community stakeholders about scenarios related to specific design choices. Our work differs from \citet{brown2019toward} in that we: 1) worked independently of a CPS agency, 2) asked whether PRMs should be used in the first place, and 3) asked open-ended questions about other technologies or non-technical changes beyond just designing algorithms for CPS agencies. Our approach can be seen as human-centered \cite{chancellor2019who,aragon2022human,chancellor2022practices}: where the humans that we center are impacted communities, not government agencies. Drawing from standpoint theory \cite{collins1997standpoint,harding2004feminist} and the Marxist roots of participatory design \cite{gregory2003scandinavian},\footnote{As Ehn describes: ``In the interest of emancipation, we deliberately made the choice of siding with workers and their organisations'' \cite{ehn1993scandinavian}.} we engaged with parents and workers who were most impacted by, but most disempowered around, decisions on data and technologies in CPS to better understand a ``view of technology \textit{from below}'' \cite{abdurahman2021body}.\footnote{While frontline CPS workers have power over families, they have little say around their working conditions nor the technologies they use \cite{cheng2022disparities,kawakami2022partnerships}.} These methodological differences may have led to novel suggestions in Sections~\ref{sec:new-uses}, \ref{sec:low-tech-alternatives}, and \ref{sec:guidelines}, which go beyond those uncovered in prior community-engaged research.

\section{Background}
\begin{figure}
     \centering
     \begin{subfigure}[b]{0.48\textwidth}
         \centering
         \includegraphics[width=\textwidth]{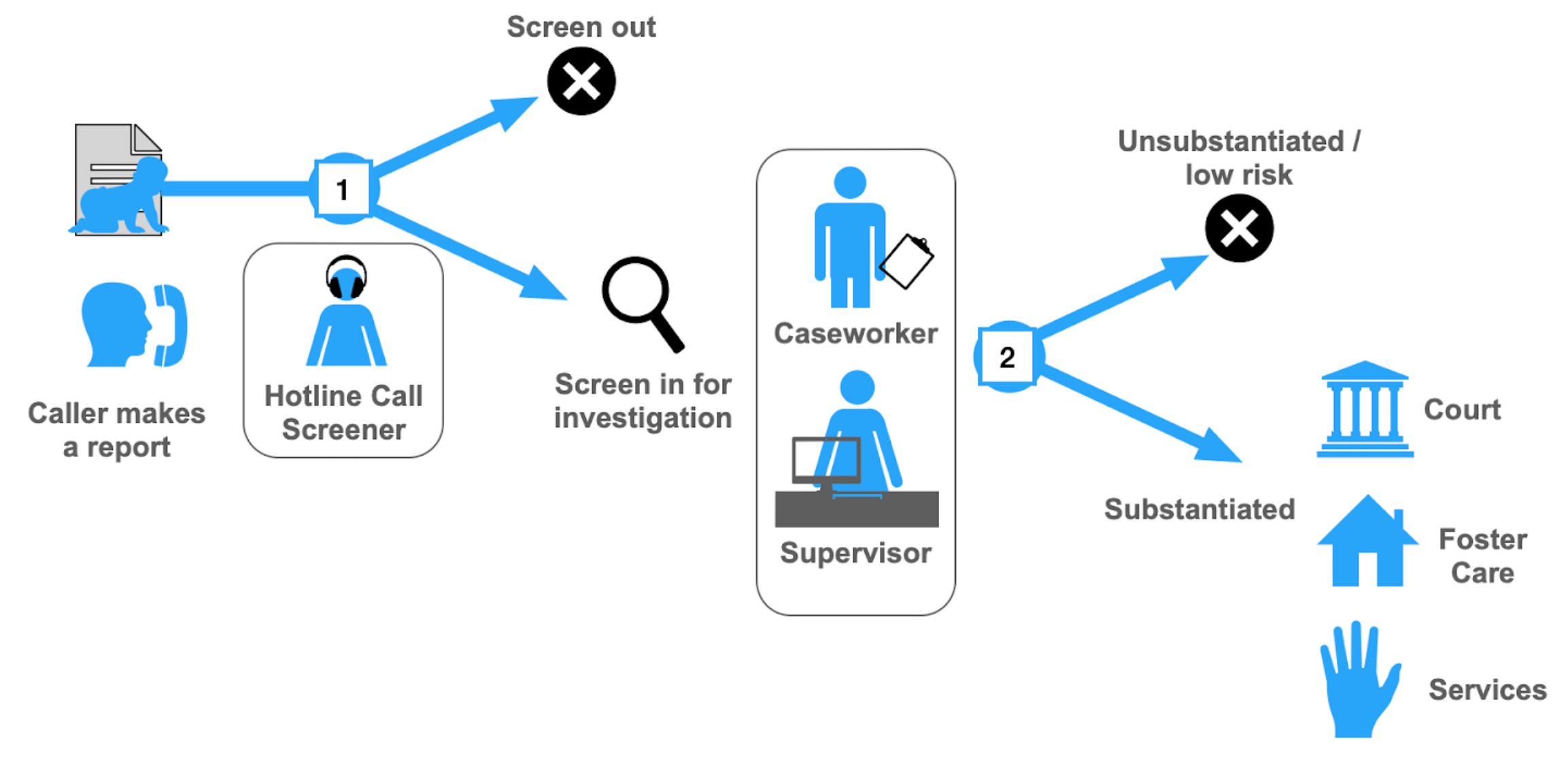}
         \caption{Front-end child welfare decision-making where PRMs are currently used, and the workers involved.}
         \label{fig:decision-making-intro}
     \end{subfigure}
     \hfill
     \begin{subfigure}[b]{0.48\textwidth}
         \centering
         \includegraphics[width=\textwidth]{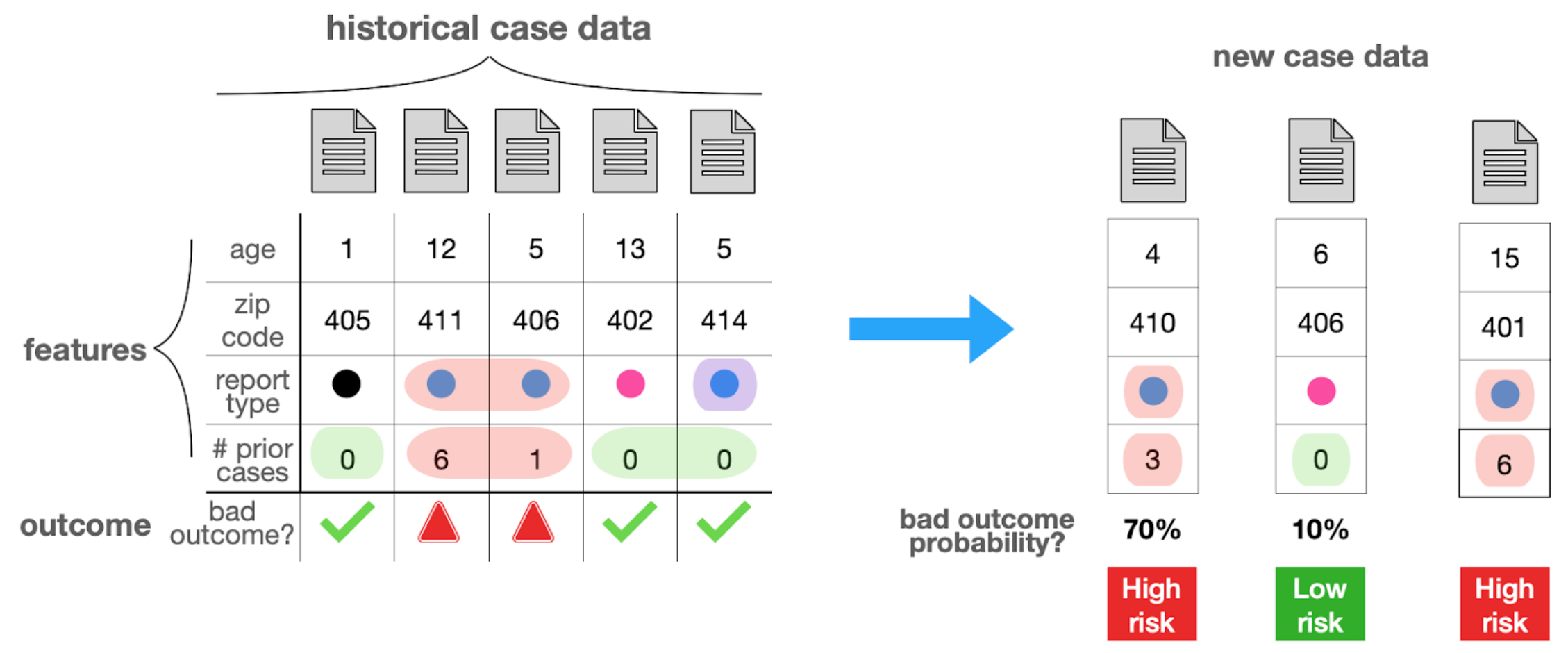}
         \caption{A simplified diagram of how PRMs are trained and used on specific cases.}
         \label{fig:data-algorithm-intro}
     \end{subfigure}
     \caption{Diagrams shown to participants in Activity 1 of the workshops to explain how current PRMs work and where they are used.}
        \label{fig:intro-explanations-overview}
        \Description{Diagrams shown to participants in Activity 1 of the workshops. There are two subfigures in this figure: one on the left and one on the right. The subfigure on the left is a flowchart diagram of the steps of front-end decisions in child welfare. The diagram contains light blue figures of first a person reporting a case to CPS, then a worker taking that call, then a caseworker and a supervisor who investigate the cases, then two arrows stemming from the caseworker saying whether the case was substantiated or not. The subfigure on the right is a simplified diagram of how PRMs are trained and used. It contains a grid with numerical and categorical variables which represents historical case data. Then there is an arrow next to that grid pointing right towards three columns of data. Each column of data represents a new case to apply the PRM to. Under each column is a red box that says High risk or a green box that says Low risk, which represents the PRM's label based on patterns in the data in that column.}
\end{figure}

Figure~\ref{fig:intro-explanations-overview} demonstrates how data-driven predictive risk models (PRMs) work and how they are used currently in child welfare. Many U.S. child welfare agencies currently use PRMs, mostly to assist workers in ``front end'' decisions, such as which families to investigate or how to investigate them \cite{eckerdrsf,vaithianathan2017,aclu2021family}. A few agencies are starting to use PRMs to allocate services to families before they are reported or to prevent foster care placement \cite{dana2019predictive,abdurahman2021calculating,hellobabyfaq}. No agencies currently use PRMs in decisions after investigation, e.g. in court; however, there are currently no regulations around how PRMs can or cannot be used. Figure~\ref{fig:data-algorithm-intro} demonstrates how a typical PRM is developed and used in CPS. Different agencies or PRMs can use different kinds of data. However, most algorithms use family demographics (excluding race) and past CPS data, e.g. about prior reports on the family \cite{chouldechova2018case,goldhaber2019impact,saxena2020human}; some use other governmental data, e.g. criminal, public health, or public benefits data \cite{vaithianathan2017}. Many PRMs are designed to predict the likelihood of some observable proxy for abuse or neglect, which are often vague and rarely observable \cite{saxena2020human}. A machine learning (ML) algorithm then uses this data to train a model (the PRM). Finally, this PRM is applied to new case data and the PRM's assessment ---interpreted as the likelihood of some proxy for abuse or neglect--- is shown to CPS workers, who use it when making decisions \cite{saxena2020human,kawakami2022partnerships}. Although no PRM is currently used to fully automate decisions, some suggest this is possible \cite{eubanks2018automating,nccpr2019racial,cheng2022disparities,De-Arteaga2020case}. Others note that automation is a spectrum: CPS agencies can pressure workers to conform to PRMs' recommendations in some cases more than others \cite{cheng2022disparities,kawakami2022partnerships}.

\label{sec:background}

\section{Methods}
\label{sec:methods}

\begin{figure}
      \centering
      \includegraphics[width=8.5cm]{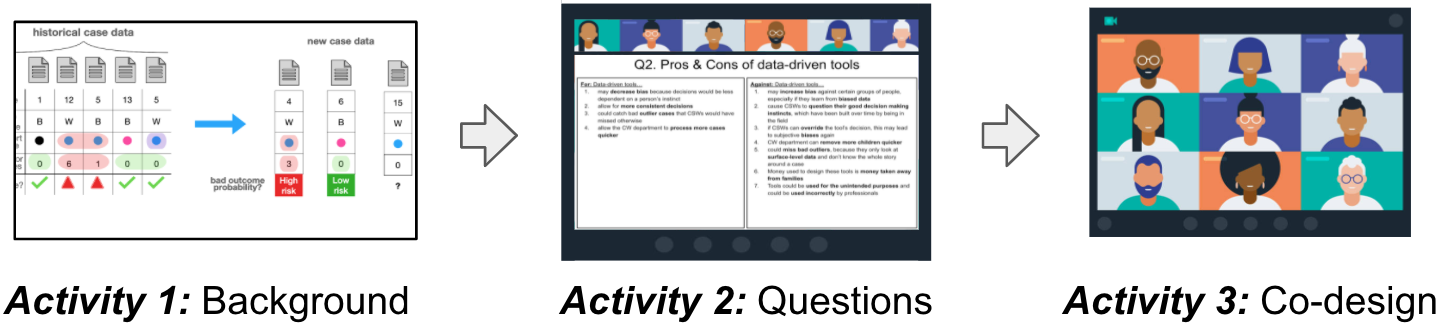}
       \caption{Outline of the study protocol, including the three workshop activities.}
      \label{fig:protocol}
      \Description{Outline of the workshop protocol, which contains three activities, each represented as a box in the figure. The three boxes are separated with right-pointing arrows. The left box is captioned ``Activity 1: Background.'' The image is the same as Figure~\ref{fig:intro-explanations-overview}, a simplified visual explanation of how PRMs are trained using historical county data and how they are applied on new case data. The middle box is captioned ``Activity 2: Questions.'' It depicts a Zoom call where workshop participants are answering questions, which are then written down on a shared Google Doc. The right box is captioned ``Activity 3: Co-design.'' It depicts a Zoom call where workshop all participants are organized in a grid talking to each other.}
  \end{figure}

Our work takes a human-centered, participatory approach to the design and use of predictive risk models (PRMs) and data-driven technologies in child welfare. We conducted 7 workshops with 35 participants total. Workshops were conducted over Zoom, each with 4 to 7 participants who were impacted by or worked in CPS.

\begin{table*}[h]
    \resizebox{16cm}{!}{%
\begin{tabular}{|p{0.45cm}|p{6cm}|p{0.35cm}|p{0.35cm}||p{0.45cm}|p{6cm}|p{0.35cm}|p{0.35cm}|}
\hline
\textbf{ID} & \centering{\textbf{CPS Personal or Job Experience}} & \textbf{Q1} & \textbf{Q2} & \textbf{ID} & \centering{\textbf{CPS Personal or Job Experience}} & \textbf{Q1} & \textbf{Q2}\\ \hline
\textbf{P1} & ED of private CPS agency; former foster youth & NA & Yes & \textbf{P19} & CPS worker & No & Yes\\ \hline

\textbf{P2} & PhD student studying public algorithms & NA & No & \textbf{P20} & Impacted parent; ED of parent advocacy group & Yes & No\\ \hline

\textbf{P3} & Attorney (family law \& ICWA) & NA & No & \textbf{P21} & Impacted parent; assistant editor & Yes & No\\\hline

\textbf{P4} & DHS licensor & Yes & Yes & \textbf{P22} & Impacted parent; parent trainer & Yes & No\\\hline

\textbf{P5} & CPS worker & NA & Yes & \textbf{P23} & Impacted parent; parent advocate & Yes & Yes\\\hline

\textbf{P6} & PhD student studying public algorithms & NA & No & \textbf{P24} & Impacted parent; parent advocate & Yes & Yes\\\hline

\textbf{P7} & CPS management & NA & Yes & \textbf{P25} & Impacted parent; parent advocate & Yes & Yes\\\hline

\textbf{P8} & Teacher (mandated reporter) & NA & No & \textbf{P26} & Impacted parent & NA & NA\\\hline

\textbf{P9} & CPS worker; lecturer & NA & Yes & \textbf{P27} & Impacted parent; parent advocate & Yes & Yes\\\hline

\textbf{P10} & Attorney & NA & No & \textbf{P28} & Impacted parent; parent advocate trainer & Yes & Yes\\\hline

\textbf{P11} & Psychologist; attorney & NA & No & \textbf{P29} & Perinatal social worker & No & Yes\\\hline

\textbf{P12} & Impacted parent & Yes & No & \textbf{P30} & CPS field director; academic faculty & Yes & Yes\\\hline

\textbf{P13} & ED of private services agency & NA & Yes & \textbf{P31} & CPS project manager & No & Yes\\\hline

\textbf{P14} & CPS administrator & Yes & Yes & \textbf{P32} & CPS worker & Yes & Yes\\\hline

\textbf{P15} & CPS worker & No & Yes & \textbf{P33} & Impacted parent & NA & NA\\\hline

\textbf{P16} & CPS administrator & No & Yes & \textbf{P34} & Impacted parent; parent advocate & Yes & Yes\\\hline

\textbf{P17} & MFT therapist; transracial adoptee & No & No & \textbf{P35} & Impacted parent; parent advocate & Yes & No\\\hline

\textbf{P18} & - & - & - & \textbf{P36} & Impacted parent; parent advocate & NA & NA\\\hline

\end{tabular}
   } % end resize
\caption{Participants' personal or job experiences with CPS. Q1 was: \textit{Have you ever been investigated by a child welfare agency?} Q2 was: \textit{Do you have child social work experience or education?} We did not explicitly ask participants about more personal experiences to avoid harmful disclosure (see Appendix~\ref{sec:voluntary-disclosure}) \cite{harrington2019deconstructing}. We omit P18's responses, since they participated in only part of a workshop.}
\label{tab:participant-experience-short}
 \Description[Participant child welfare experience]{Participant child welfare experience}
\end{table*}

\xhdr{Recruitment \& Demographics.}
Our participants were mostly impacted parents and caseworkers, plus a few private service providers, psychologists, attorneys, students, one former foster youth, and one adoptee. Table~\ref{tab:participant-experience-short} describes participants’ personal and job experiences in CPS. See Table~\ref{tab:participant-demographics} in Appendix~\ref{sec:demographics-responses} for participants' demographics. The majority of participants were Black and/or Latina women in New York or California, however there was a mix of racial/ethnic backgrounds, genders, and locations represented. 14 participants said they were impacted parents. 20 said they worked for a CPS agency or had an education in child social work --- of these, at least 8 worked in public agencies. Only 2 participants had significant technical knowledge about PRMs. Children under 18 were excluded. We recruited 23 participants through an online recruitment form distributed via email using a snowball sampling approach. We reached out to multiple existing contacts who work in CPS or teach in schools of social work in the U.S to distribute our recruitment form. We also recruited 13 participants through an existing contact in an organization for impacted parents in the northeastern region of the U.S. This agency also trained parent advocates, which is likely why many parents in our study also said they worked in CPS. Many participants had a mix of CPS experiences, e.g. workers who had been investigated. Thus, each participant reflects deep knowledge of multiple aspects of CPS and impacted communities.

\xhdr{Protocol.}
See Figure~\ref{fig:protocol} for an illustration of our study protocol. Participants were given an almost identical short survey before and after the workshop to gauge their opinions on CPS and PRMs. See Appendix~\ref{sec:survey} for a full list of survey questions and a description of responses.\footnote{As discussed in Appendix~\ref{sec:survey}, post-survey responses showed that the workshops did not significantly change participants' perspectives.} Workshops were semistructured, starting with 10 minutes of background on PRMs (similar to Section~\ref{sec:background}) including showing Figures~\ref{fig:decision-making-intro} and \ref{fig:data-algorithm-intro}, followed by a 60-minute conversation led by three questions about CPS and PRMs (see below), ending with a 20-minute design activity to elicit ideas about how to use and design PRMs, and how to improve child welfare beyond PRMs. Throughout each of these study activities, we tried to present information about PRMs and avoid value judgments of participants' responses to not sway participants.

In the Questions phase (Activity 2) of the workshop, we asked participants three questions to center conversations:
\begin{enumerate}
    \item What do you think are the goals or outcomes of an ideal system for protecting children?
    \item What are some pros and cons of PRMs?
    \item How should workers and interventions look in an ideal system for protecting children?
\end{enumerate}

Although the workshops were centered around PRMs, the first and third questions did not specifically mention PRMs in order to give space for participants to bring up comments or concerns about CPS in general. For each of these questions, we shared a document with participants to add their comments to. Our team of facilitators also took notes in real-time. We did not erase the documents between workshops, so that participants in later workshops could comment on past participants' thoughts.

In the Co-design activity (Activity 3), we asked participants to write down at least 4 ideas to change PRMs or CPS. Then, we asked each participant to share 2 of their ideas and write those on a shared document. Finally, we asked participants whether they agreed or disagreed with other participants' ideas, and asked the group to make one collective list of ideas (without mandating consensus). Our design activity was based on Crazy Eights \cite{knapp2016sprint,designkit2021}. Though there may be drawbacks to these kinds of open-ended design activities \cite{harrington2019deconstructing}, we draw inspiration from abolitionists in ``imagining a safer world'' for Black and other minoritized people \cite{roberts2022torn}.

\xhdr{Ethics \& Institutional Review.}
To minimize the risk of unintended harms to participants, we consulted domain experts and impacted parents when designing our study \cite{pierre2021getting,harrington2019deconstructing}. Two workshops included only impacted parents to reduce the risk of conflicts or power imbalances with other kinds of stakeholders. We also worked with leaders of the parent organization who helped with recruiting assist in facilitating these two workshops. We did not ask participants to disclose personal experiences with CPS (besides whether they had been investigated), due to potential harms of such disclosures \cite{harrington2019deconstructing}. However, as a result, participants may have had additional relevant personal experiences that they did not disclose to us. This study, including all questions, study materials, and recruitment methods, was approved by the Institutional Review Board of Carnegie Mellon University.

\xhdr{Qualitative Analysis.}
We transcribed all 10.5 hours of online workshop recordings into text, then used thematic analysis~\cite{braun2006thematic} to analyze our data. We conducted open coding on the data, generating over 1000 codes. We performed an affinity mapping process, comparing and clustering alike codes, then identified themes that emerged from this affinity mapping. Examples of themes include: problems with diagnostic checklists, labels and stigmatization, and decisions not to use PRMs for. In Section~\ref{sec:results}, we present a subset of these themes which are most relevant to FAccT readers, leaving out some themes specific to child welfare which did not pertain to PRMs nor future design work.

\xhdr{Positionality.}
Most of the authors of this paper are academic ML and HCI researchers, white or Asian, and have little personal CPS experience (although one author is also Black and Latina and works in CPS). Our participants are mostly frontline caseworkers or Black and Latina mothers who have been in the system. The lead author, who ran all workshops, is a white man, which may have influenced participants' responses \cite{ogbonnaya-ogburu2020critical}. We anonymize participants' responses so that they could speak freely (especially workers who may be retaliated against). At the same time, we acknowledge that this may mean we quote and get academic credit for the ideas of participants with different lived experiences than most of us. Yet, we also consider ``the researcher as an active participant throughout the research process'' \cite{copeland2021only}. We see this work as depicting a conversation between us ``socially-minded'' technological researchers and our participants, who are impacted by the technologies that our field has (or we have) created.

\xhdr{Limitations.} 
\lsdelete{Child welfare agency leadership and designers of PRMs were absent from our pool of participants (although two agency administrators participated in our study). Some participants (even impacted parents who opposed child welfare and PRMs) said they would have liked to talk more directly with proponents of PRMs. For example, P35 said, \textit{``I would like to have had [CPS] upper management in on how they see the [PRMs] and how they think [they] will be used.''} Future work in this direction may include studies with child welfare agency leadership and algorithm designers to better understand why they want to use these algorithms, or workshops which include both impacted communities and leadership or designers. } One limitation of our work is that we recruited few foster youth and adoptees, who may have differing views from the mostly parents and workers we spoke with. Another is that our study was not geographically restricted. Because CPS differs by location, our participants' responses do not necessarily reflect a specific community (nor do we claim them to). Future work may include qualitative studies focused on former foster youth and adoptees, or focused on a specific locale (e.g. one county or city).

\section{Results}
\label{sec:results}

In this section, we present prominent themes that emerged from the workshops, which we believe to be most interesting to FAccT readers. See Table~\ref{tab:suggestion-summary} for a summary of suggestions. Section~\ref{sec:participant-concerns} outlines participants' concerns with PRMs, which many viewed as exacerbating existing problems in CPS. In Section~\ref{sec:new-uses}, participants offer suggestions for new work that researchers can do for impacted communities, beyond creating PRMs for CPS agencies. Section~\ref{sec:guidelines} includes suggestions on how to mitigate potential harms caused by PRMs if they continue to be used. Section~\ref{sec:low-tech-alternatives} offers no-tech or low-tech alternatives which may better address many of the problems that have motivated the use of PRMs.

\begin{table*}[h]
{\renewcommand{\arraystretch}{1.2}
\begin{tabular}{|p{3cm}|p{11.5cm}|}
\hline
\textbf{Section} & \textbf{Participants' Suggestion}\\ \hline
\multirow{2}{3cm}{Harms of PRMs (Section~\ref{sec:participant-concerns})} & PRMs reinforce CPS' punishment, undersupport, disempowerment of families \\ \cline{2-2}
& PRMs perpetuate existing biases and racism in CPS \\ \hline
\multirow{4}{3cm}{Data \& design beyond PRMs (Section~\ref{sec:new-uses})} & Researchers \& designers should work in solidarity with impacted families (to oppose CPS) \\ \cline{2-2}
& Use data to evaluate CPS, workers, reporters, interventions, etc \\ \cline{2-2}
& Technology to recommend mandated reporters when not to report \& where to reroute calls \\ \cline{2-2}
& Use PRMs to allocate resources (but not if this expands surveillance) \\ \hline
\multirow{6}{3cm}{Mitigating PRM harms (Section~\ref{sec:guidelines})} & Strict regulations on how data \& PRMs can \& cannot be used \\ \cline{2-2}
& Regular evaluation of PRMs before \& after deployment, especially on racial biases \\ \cline{2-2}
& Give impacted families more control over CPS policy, data, \& technology decisions \\ \cline{2-2}
& Include data on CPS, workers, reporters, interventions, etc in PRMs \\ \cline{2-2}
& Do not use demographics nor zip codes in PRMs \\ \cline{2-2}
& PRMs (and CPS more broadly) should focus on strengths, rather than deficits\\ \cline{2-2}
& Do not fully automate CPS decisions \\ \hline
\multirow{5}{3cm}{Low- \& no-tech alternatives to PRMs (Section~\ref{sec:low-tech-alternatives})} & Improve hiring, training, working conditions, \& team-based decision-making \\ \cline{2-2}
& Make policy \& legislative changes to address systemic harms \\ \cline{2-2}
& Give money directly to families instead of spending on CPS or PRMs \\ \cline{2-2}
& (Maybe) use diagnostic checklists \& practice models instead of PRMs \\ \cline{2-2}
& Abolish the child welfare system \\ \hline
\end{tabular}
} % end stretch
\caption{Summary of participants' suggestions presented in Section~\ref{sec:results}.}
\label{tab:suggestion-summary}
\end{table*}

\subsection{Concerns that PRMs reinforce systemic problems in CPS}
\label{sec:participant-concerns} 19 of 32 participants who responded to our survey disagreed that current PRMs would lead to better outcomes in CPS; only 5 agreed (8 were neutral). Personal experiences led many participants to hold negative views of CPS, e.g. P1 who explained her views simply by: \textit{``31 years working in the system.''} Like \citet{brown2019toward}, our participants disliked PRMs due to ``system-level concerns;'' yet, our participants gave more pointed criticisms.

\textbf{Participants disliked PRMs for further entrenching CPS in what they saw as punishment, undersupport, and disempowerment.} Participants (both parents and workers) said that CPS often punishes families instead of supporting them. P24, a parent, said, \textit{``so many people have been treated badly... when they were on a good foot, but because they don't have enough support, certain things got out of hand, and they wasn't given the opportunity to pick up the pieces... They just automatically get scolded and child removed.''} P9, a caseworker, echoed this, using the disparate treatment of foster families versus original families as an example: \textit{``we punish the [original] parents for not doing all the right things''} while \textit{``our foster family agencies have a plethora of resources and support and funding to ensure that that child's needs are met.''}\footnote{For example, P9 said, \textit{``on welfare, a mother of one or a few children is only going to receive between \$300 to \$400 a month, and that is now in the state of California capped out... For a foster parent... the least amount that I've seen in my county is \$1,000 a month.''}}

Participants said current PRMs would not help support families. P12 said that CPS' \textit{``goal is really to support families, and I just don't think this tool plays any role in actually supporting families.''} Rather, participants said PRMs widen surveillance by encouraging CPS to process more cases and intervene more (P1,P13,P20,P30), getting more families involved in CPS (P7,P11,P12,P13), and getting more families stuck in the system for too long (P7,P14,P27,P29,P33). P12 said that she worried PRMs would cause \textit{``more monitoring, more surveillance, more intervention in Black and Brown and poor communities.''} P27 said, \textit{``I don't trust the algorithm, because it's... been set up to just surveil Brown and Blacks.''} McMillan explains: ``It's surveillance:... Coming into someone's home, checking their drawers, cabinets, and strip searching their children, how is that support?'' \cite{webeimagining2020mothers}. A number of our participants said this exact scenario happened to them or happens regularly to people they work with, e.g. P26 said CPS came to \textit{``strip my kid butt naked and go through my cabinets and uproar and turn my whole house upside down.''} P12 described calling a domestic violence hotline for help, and instead getting investigated by CPS and having her child removed. PRMs claim to improve efficiency: Participants said this could be helpful if it got families out of the system quicker, but harmful if it got more families investigated (P7,P11,P12,P13). Participants saw similarities between PRMs in CPS and the criminal system (e.g. \cite{propublica2016compas,albright2019judge,stevenson2021algorithmic,stevenson2018assessing}) and the use of criminal data in PRMs in CPS (e.g. \cite{vaithianathan2017}) as further solidifying CPS as a carceral institution (P33,P35,P36). P33 worried PRMs would embolden CPS workers and police, allowing them to act like \textit{``attack dogs''} on families with high risk scores. P12 said PRMs would add an \textit{``extra layer''} for parents to fight through: \textit{``not only are you fighting the... agency, now you're gonna have to fight this computer system.''} Workers also worried about an extra layer of blame if they disagreed with a PRM, reinforcing a ``Cover Your Ass'' mentality (P3,P7,P9,P19).\footnote{P19 said, \textit{``If we still do have a child fatality... then it's another [reason] to be like `Well, you had this tool and this tool told you that this family needed X, Y, Z.'''}}\footnote{This is similar to treating workers in the loop as ``moral crumple zones'' \cite{elish2019crumple}.} Participants said PRMs reinforce caseworkers' power over families and their role as gatekeepers. P29 said, \textit{``the power holder is the caseworker that's inputting the information and so it's already starting from a standpoint of they're the end-all be-all.''} Finally, participants said PRMs allow designers and CPS leadership to control on-the-ground decisions and justify harms. P12 said, \textit{``Computers don't make decisions; people make decisions and program the computers to carry out those decisions. So we're not going to turn around and say, `Wow. Oh, it's the computer that's creating this decision and this is why 80\% of the children who go into foster care are from the Black communities.'''}

\textbf{Participants said PRMs perpetuate existing biases and racism in CPS.} 26 participants said they did not trust CPS to make unbiased decisions; only 1 participant said they did. P12, whose child was placed in foster care, said, \textit{``I've been through the system, I know how harmful it is and how racist it is and how destructive it is to Black and Brown and marginalized communities and even poor people.''} Participants thought PRMs would not address the most prominent causes of biases based on race or class, such as laws and policies which justify differential treatment of poor and Black families within CPS, or biased reporting outside CPS. Participants also thought that PRMs would not eliminate workers' biases because they still allowed for worker discretion (P1,P5,P6,P7,P9,P15,P16),\footnote{As we discuss further in Section~\ref{sec:guidelines}, many also did not want to fully automate decisions with algorithms (P2,P6,P10,P14,P17).} and would even exacerbate racial biases because of biased or ``dirty'' data. \Citet{brown2019toward} heard similar worries about biased workers and data; yet our participants go further, saying that PRMs reinforce racism and classism in CPS. Participants said CPS stigmatizes poverty (P1,P2,P5,P10,P14,P33,P36) and PRMs which use governmental records and demographics justify and exacerbate this (P1,P2,P5,P33). Overall, most participants suggested that PRMs were at best ineffectual, and at worst counterproductive, at mitigating existing discrimination and disparities based on race and class (see \cite{dettlaff2020racial}). Some participants thought PRMs would perpetuate or exacerbate other existing biases, e.g. against former foster youth (P1,P5,P36) or people with mental illnesses (P36).

There were some exceptions to these overall sentiments. Though, even those who liked PRMs said they might reduce individual workers' biases and improve decision-making, but they would not address systemic issues. P4 said larger reforms were needed to address systemic discrimination, but that these changes would not happen overnight. \textit{``In the meantime,''} P4 thought PRMs could help day-to-day decisions now, especially if they used the \textit{``right data''}: \textit{``If you put the correct data points in... maybe we can take some of that subjective bias out of it.''} A few other participants echoed similar sentiments about incremental benefits of PRMs coupled with systemic changes (P13,P14,P31). This sentiment of PRMs helping ``in the meantime'' has been echoed by proponents of PRMs, including CPS agencies defending their use \cite{DHSResponse}. Beyond these exceptions, most participants saw current PRMs in CPS as exacerbating what they saw as CPS' tendency to punish instead of support families, particularly poor, Black, Brown, and Indigenous ones.

\subsection{Beyond PRMs: New directions to work in solidarity with impacted communities}
\label{sec:new-uses}
Although most participants opposed current PRMs, many gave constructive suggestions on how researchers and designers can use data and technologies to support impacted communities, beyond just designing PRMs for CPS agencies.

\textbf{Participants suggested that researchers and designers should work in solidarity with impacted families and communities to use data to oppose CPS} (P19,P24,P25,P36). For a number of participants, the desire for researchers to work with communities manifested through suspicion that us authors were working with CPS agencies or did not have communities' interests at heart.\footnote{To reiterate, we told participants that our study was being conducted and funded independently from any agency.} P11 speculated that our study was being conducted by the \textit{``inventors of [PRMs]''} in order to \textit{``anticipate... the objections... of potentially skeptical people [so that] the sponsors will be [better equipped]... to resist the objections''} in order \textit{``to further develop their tools and sell them, and thus become prominent in their academic fields, or make money, or both.''} In another workshop, P24 asked, \textit{``What is the point of all this data-driven focus mess?''} then asked the lead author to consider whether they were doing this work to publish a paper and further their academic career or whether it was work which could actually benefit families harmed by CPS.\footnote{The lead author especially appreciates this personal confrontation to push his thinking and work in the right direction.} Given that most prior work on CPS in ML and HCI has been conducted to help develop algorithms to assess families or in partnership with CPS agencies \cite{brown2019toward,cheng2022disparities,kawakami2022partnerships,saxena2020participatory,saxena2021framework}, these suspicions seem justified. Instead, participants suggested specific ways researchers could better work in solidarity with communities. For example, participants suggested using data about families who have successfully fought CPS to produce strategies and suggestions for other impacted families to do the same (P13,P20,P24). Others suggested using data to help parent advocates verify or disprove negative and/or erroneous claims that CPS agencies make about parents (P25,P36).

\textbf{Participants suggested using data to evaluate the child welfare system and the people who work in it}, including reporters of alleged abuse, foster parents and homes, CPS workers, agencies, interventions, services, etc (P1,P2,P4,P6,P9,P12,P25,P28,P29,P30,P31,P33). Participants said administrative data collected on families reflect more on CPS and other governmental systems than they do on individual parents (P1,P3,P4,P12). P1 said, \textit{``if you've had 6 open cases, that means [CPS has] had 6 times where we weren't helpful to a family. It's measuring the system... It doesn't tell us anything about the people.''} Participants thought that data and data-driven tools (such as PRMs) should be used to assess harms caused by CPS and help communities push for change (P1,P4,P10,P23,P28). P10 said, \textit{``it doesn't make sense at all to me, why high or low risk is even what anyone thinks is being predicted... [PRMs] could just as easily be measuring the extent of racism, the extent of surveillance.''} This hearkens back to Roberts' \cite{roberts2002shattered} call to ``measure the extent of community damage caused by the child welfare system.'' For example, data-driven tools could be used to evaluate CPS workers, like they have been used on other street-level bureaucrats \cite{carton2016identifying}.

\textbf{Participants suggested designing an algorithm for mandated reporters to recommend whether and where to make a report} (P14,P19). P19 said such an algorithm should address the following questions: \textit{``Is this something I should make a call on? Is this something I should reach out to a prevention agency or agency that could possibly service the family prior to just calling it into [CPS]?''} P14 and P19 said the goal here is to reduce the number of families in the system, either by recommending not to report or rerouting calls somewhere else.

\textbf{Participants suggested using PRMs to allocate resources, but some worried this would expand surveillance and stigmatization} (P1,P2,P4,P14,P31). Although many participants said PRMs should not be used for coercive interventions, e.g. investigations or home removals, some suggested using PRMs to connect families to resources and services. P2 said, \textit{``what I would want to see in the future is using these tools to decide on resource allocation, like who should have priority for access to services; instead of starting an investigation, more framing it from a more positive and supportive side.''} Specifically, participants wanted more direct assistance to help with childcare or alleviate poverty, which many viewed as a common root cause of neglect and abuse (which is backed by prior work \cite{drake2014poverty}). P7 said, \textit{``the goal would be to... have finances available to support families in need as a preventative measure, or housing, or employment, or... medical services''} or even something like \textit{``Supernanny [to] go into homes and be there to help the family.''} Beyond individual assistance, some participants suggested community- or neighborhood-based approaches \cite{casey2000neighborhood,roberts2005community}. P1 suggested to \textit{``use data to find the top 3 zip codes where child protection is involved and get some of our local Fortune 500 companies to create living wage jobs in those zip codes.''}

However, participants also worried that expanding services provided by CPS or connected to CPS through mandated reporters would expand surveillance and place a stigma on families.\footnote{Some participants advocated for getting rid of anonymous (or all) mandated reporting to decrease the chances that assistance would lead to CPS intervention.} P15, a caseworker, said, \textit{``those who... have more contact with systems... are the ones who get reported on constantly.''} P1, a private CPS worker, said, \textit{``people can't ask for help without a report.''} P33, a parent, said, \textit{``[PRMs] put a stigma on people themselves... You know, it's not like anybody's saying, `Well, I want my significant other to run out and leave me with the child by myself and I struggle, so I had to get on welfare.' ... Basically to survive, I get a stigma.''}\footnote{P14 suggested that this may be a problem of semantics, suggesting that replacing `high risk' with `high need' might make communities more comfortable. However, other participants said they would be uncomfortable with any label from a PRM.} P20, a parent, said this leads \textit{``communities [to] hide in their struggles [rather] than say they need support or reach out for needed resources.''} Prior work describes this tension where families want more supportive resources but fear more CPS intervention \cite{roberts2008paradox,burton2021toward,roberts2022torn}. Recent work shares our participants' fear that PRMs used to allocate services will ``[sweep] into the carceral net low-risk individuals who previously would not have been on the  government's punitive radar at all'' \cite{roberts2019digitizing,abdurahman2021calculating}. Empirical work suggests that PRMs which use data on public services may lead to over-surveillance of Black families \cite{cheng2022disparities}. Some participants (P14, P19) worried about using PRMs for ``preventive services,'' which are services CPS agencies offer to prevent child maltreatment or future CPS involvement \cite{antle2009prevention,casey2021prevention,testa2020evolution,waldfogel2009prevention}. Recent work \cite{abdurahman2021calculating} suggests that PRMs will increasingly be used for preventive services, due to funding from the newly-enacted Family First Prevention Services Act (FFPSA), early examples in New York and Pittsburgh to look to \cite{hellobabyfaq,abdurahman2021calculating,dana2019predictive}, and to avoid criticism like that of PRMs used for screening or investigations \cite{eubanks2018automating}. P14, an administrator, confirmed their agency is doing exactly this: \textit{``The [FFPSA] is... requiring a lot more evidence-based preventive services... One of the things that [our agency is] looking at is `What about primary or secondary prevention?' In Allegheny County, they have another... preventive risk modeling tool called Hello Baby''} \cite{hellobabyfaq}.

\lsdelete{
\textbf{P31 proposed using data to better track progress and outcomes for families.} P31 said most CPS agencies currently measure family progress by tracking which services they are provided and which they complete. However, P31 said they should use data (which P31 said agencies will likely already have) to track better outcomes, e.g. \textit{``Are they making appointments? Are they doing this? Are they being responsible? As opposed to [an] array of services and it's a checklist and it feels like you're just going through the motions.''} P30 agreed and said that tracking progress using more granular, personalized data would be a way for agencies to better calibrate services to families, e.g. to follow families as they progress slower or faster, instead of judging families based on standardized time frames for service completion.
}

\subsection{Guidelines for mitigating harms of PRMs}
\label{sec:guidelines}
As stated in Section~\ref{sec:participant-concerns}, most participants opposed current PRMs. Yet, many said that if these tools were to continue being used, they would like more guidelines around their use and design to reduce harms.

\textbf{Participants wanted stricter rules on how data and PRMs can and cannot be used}, so that data collected, or tools designed, for one purpose do not end up being used for another purpose (P2,P13,P33). P2 said they \textit{``would like to see some sort of policies to be put in place that would prevent tools like this being misused in the future... [and] really strict guidelines about how we can use these tools.''} For example, local governments could implement legislation like Community Control Over Police Surveillance (CCOPS), which requires elected representatives to approve any government data or surveillance technologies (including PRMs) \cite{ccops}. Some participants said PRMs should not be used for placement decisions (P10,P12,P26,P33) nor day-to-day decisions in general (P12,P17).

\textbf{Participants said PRMs should be evaluated before and regularly after deployment} (P4,P14,P30,P35). One big reason agencies have said they use PRMs is to mitigate workers' biases and address racial disparities in the system. Our participants suggested evaluating PRMs on whether they actually do this. Recent work suggests this, as well \cite{drake2020practical,green2021flaws}. See, for example, prior work auditing PRMs \cite{goldhaber2019impact,cheng2022disparities}. However, P6 thought that evaluating whether algorithms help or harm may be difficult, especially if overall group effects such as racial disparities are improved, but individual families are harmed more. Future work on auditing algorithms should clarify how best to measure group and individual impacts.

\textbf{Participants wanted more impacted families involved in CPS policy and technology decisions.} Participants recommended that involving communities to make decisions and set policies can help mitigate biases at the unit- or agency-level (P5,P10,P14,P17,P24,- P26,P29,P30). P14, an administrator, said, \textit{``I think that families are the experts of their own, particularly even youth.''} P23, a parent, said, \textit{``we have to be part of the language that's controlling and setting the laws and that's... happening at every level of engagement for our families.''} \citet{roberts2002shattered} argues to shift control of CPS to Black families, specifically. Participants said families should be more involved around how new technologies are used and created (P1,P2,P5,P7,P10,P13,P14,P15,P22,P29). P14 said, \textit{``if you're developing anything, [it] needs to be community-led and... family led.''} P29 said, \textit{``those that are creating [PRMs] should also be diverse and really reflect the communities that will be impacted by it, so that they're thinking... intentionally.''} P22 said that PRMs might help make more equitable decisions \textit{``if they understood parents more.''}

\textbf{Participants said PRMs evaluating families should at least include data on CPS, reporters, workers, foster parents or homes, agencies, interventions, services, etc} (P1,P4,P7,P10,P30). P1 said, \textit{``I don't know how you create these tools to measure the right thing if the data that goes in doesn't include specifically who the child protection social worker is, what intervention they received, and at what dosage;... unless you're measuring the other half of the equation, it's hard for me to imagine that you can get a good assessment.''} Prior work also suggests including intervention data in PRMs \cite{coston2020counterfactual}. This is feasible, since data is already collected on all parts of the system except for anonymous reporters. However, JMacForFamilies is campaigning for NY State Senate Bill S7326 to require data collection on reporters in New York \cite{jmacforfamilies}.

\textbf{Participants said PRMs (and CPS more broadly) should focus on strengths, rather than deficits of families} (P1,P3,P4,P12,-
P13,P16,P17,P29,P31,P35). By focusing primarily on risk factors and predicting negative outcomes, P29 worried PRMs put families \textit{``at a deficit''}. P35 worried that PRMs do not adequately \textit{``take into account the... things [families] may have done or are doing to keep [their] child safe.''} Instead, P13 said PRMs should predict \textit{``strengths and success.''} More broadly in CPS, P14 said, \textit{``the narrative that we think about families needs to shift, as well, to one of a strength-based... interaction.''} This sentiment is echoed in prior work, as well \cite{saxena2020human,holtenmoller2020shifting}.

\textbf{Participants said PRMs should not use 
demographics nor zip codes} (P3,P12,P15,P26,
P27,P28,P29,P30,P33,P36). Participants worried PRMs using zip codes and demographics (which are correlated with race and class) would justify discrimination of poor and Black families (P1,P2,P5,P33). Participants said using demographics was not new to CPS. For example, P12 said, \textit{``right now without using data analytics, they’re still looking at your age, they’re still looking at your zip code.''} Yet, PRMs justify this practice. Participants said using zip codes and demographics was disciminatory because these factors were irrelevant to parental (un)fitness. P28 said, \textit{``it is unfair to say `because I live in this neighborhood, that must mean I'm a shitty parent'... It's unfair... to say `6 out of 10 of my neighbors had had [a CPS] case, so it's most likely I'm gonna have [a CPS] case'.''} P26 said it \textit{``makes no sense''} to use \textit{``your demographics, or past somebody else's history to determine whether you're a fit parent... because life is unpredictable.''} Prior work argues people are unpredictable \cite{birhane2021impossibility}.

\textbf{Participants said PRMs should not automate CPS decisions} (P2,P6,P10,P14,P17). P17 said, \textit{``[full automation is] too much power, it's too much impact, and 99.9\% of the time, [the PRM] fails.''} P6 pointed out a tension between automation versus worker bias: \textit{``I don't... believe that we should just hand the entire decision-making process over to a tool... [But] if we allow a caseworker to override the tool's guidance... then is that just sort of a form of bias in itself?''}\footnote{Automation is not all or nothing: forms of `soft automation' include agencies mandating or pressuring workers to follow PRMs \cite{cheng2022disparities,kawakami2022partnerships}.} Prior work has also grappled with this tension: some argue humans in the loop often make biased decisions \cite{albright2019judge,green2021flaws}. Others argue more automation can worsen disparities and decision quality \cite{De-Arteaga2020case,eubanks2018automating,cheng2022disparities}.

\subsection{No-tech and Low-tech Alternatives to PRMs}
\label{sec:low-tech-alternatives}
Participants suggested changes they thought would better address many of the problems motivating the use of PRMs, particularly which do not require AI-based technology (low-tech) or require no technology at all (no-tech).\footnote{We borrow ``low-tech'' and ``no-tech'' from \citet{baumer2011implication}.}

\textbf{Participants suggested improving hiring, training, working conditions, and team-based decision-making instead of PRMs.} \lsdelete{Overall, participants thought that improving these factors would help workers make good, consistent, unbiased decisions more than a PRM would. } First, participants said improved hiring practices would improve decision-making and alleviate biases, instead of using PRMs (P17,P19,P23,P24,P26,P32). Some said agencies should be more selective about who they hire; P24, a parent, said, \textit{``they need to stop hiring workers who just come out of college that don't have no children or have real life experience.''} Some also thought hiring more diverse workers could decrease racial biases (P29,P31,P32).\footnote{Though, some prior work argues that diverse or ``culturally-sensitive'' workers do not resolve racialized harms or discrimination \cite{roberts2002shattered}.} Second, participants said CPS agencies should improve supervision, especially of young or inexperienced workers (which is common in CPS \cite{edwards2020characteristics}) (P10,P16). Third, participants said team-based decision-making (especially diverse teams) could alleviate workers' individual biases (P7,P9,P15,P17,P31,P32). Fourth, participants said agencies should improve worker training (P10,P14,P16,P17,P19,P24). Finally, participants said agencies should improve working conditions, such as giving workers more time to make decisions, reducing caseloads, and increasing pay (P16,P17,P26,P32). This is important, since high case volumes have been a motivation for PRM use.\footnote{For example, Emily Putnam-Hornstein said Allegheny County created the Family Screening Tool \cite{vaithianathan2017} because they ``were fielding significant volumes of calls... and they were trying to figure out whether they could use data'' to address this \cite{netflix2020gabrielfernandez}.} Participants also suggested smaller caseloads would reduce turnover, which would help retain workers who were hired and trained properly and reduce the number of new, inexperienced workers. P16, a retired administrator, said, \textit{``I have always found that workers that were well-supported ---and whatever that means to them, not as the administration defines--- can be very helpful in the longevity and the decreasing of turnover.''} P19 said PRMs should be unnecessary: \textit{``if you're a good social worker, you already know which one of your cases are more high risk and how to prioritize those cases.''} P26, a parent, said, \textit{``[CPS] staff needs to be trained better, paid better, and maybe if they had happy workers, they care about their job and what they do.''}

\textbf{Participants wanted policy and legislative changes instead of PRMs} (P3,P4,P9,P19,P24,P26,P29). Participants said a lot of systemic biases in CPS are caused by laws and policies. For example, P20 said many old laws \textit{``harm families, or target low-income Brown and Black families.''} In order to address systemic biases, participants recommended changing these laws. Participants suggested changing mandated reporting laws (P13,P19,P24). P26 suggested repealing laws and policies, like the Adoption and Safe Families Act (ASFA) \cite{adler2001asfa}. These echo growing movements to repeal ASFA \cite{repealasfa} and change mandated reporting laws \cite{jmacforfamilies}. P4 also said they want new funded mandates to get resources to communities and address systemic problems.

\textbf{Participants suggested giving money directly to communities instead of spending it on CPS services or PRMs} (P1,P2,P7,- P11,P12,P13,P24,P26,P33). P12 said, \textit{``the people making [PRMs]... financially benefit,... where this money could be set to pay for housing and other basic needs.''} For example, PRM developers in Allegheny County were paid over \$1 million \cite{afstfaq}. Beyond development costs, participants also noted ongoing training and maintenance costs. P13 said, \textit{``How much it's gonna cost to train... the child protection workers [to use PRMs]... is also money that's being taken away from families.''} Allegheny County also hired specific employees (``Data Entry Specialists'') to help with data entry for their PRM \cite{vaithianathan2017}.

\textbf{Participants proposed using diagnostic checklists and practice models instead of PRMs, but others said these low-tech tools had their own problems.} Some suggested using diagnostic checklists (e.g. SDM \cite{sdm}) or practice models (e.g. SofS, SOP \cite{turnell1997aspiring}) instead of, or alongside, PRMs to alleviate workers' individual biases and improve decision-making (P1,P2,P17,P19). P1 suggested \textit{``integrating things like Signs of Safety. There are practice models... that help [workers] explore some very concrete, specific questions that help force them not to just make decisions based on their own hunch.''} However, other participants said these low-tech tools had built-in biases (see \cite{saxena2022how}) and workers frequently manipulate them (against their training) to produce any desired output (P1,P5,P6,P7,P9,P15,P16).\footnote{P5 said, \textit{``I was trained... using Signs of Safety and SOP and saying, `Well I may see this risk but I'm seeing protective factors that I think mitigate that, so I'm going to override [SDM] and not do that.'... But in practice, that's bullshit. I will override to make a 10-day an IR [Investigative Response] all the time.''}} Some said diagnostic checklists could be used better if workers were better trained and held accountable to follow the training (P5,P10,P14). Other participants thought tools should spur thought and nudge workers towards good decisions, not predict bad outcomes or give specific recommendations. P17 praised the Columbia-Suicide Severity Rating Scale (C-SSRS) \cite{posner2008columbiasuicide,posner2011columbiasuicide}: \textit{``There are some yes or no answers and it's not about ‘Oh, I want to get this kid 5150ed,'\footnotemark it's seeing what is the next step with a *thought*. So if I have information, then I use my *brain*, if I'm a human behind it. And I'm not the only one making this decision: I'm with a team.''} Finally, some participants saw PRMs as a repeat of diagnostic checklists. When presented with a list of pros and cons to PRMs, P16, a retired administrator, said, \textit{``all these things you have up here are just the same sort of precursor work they did for [SDM] before it came into play. It's no different... and in child welfare things tend to cycle back, probably, you know, decade on, decade off, decade on. So I'm just very curious about what's bringing this up again.''}
	
\footnotetext{5150 is involuntary hospitalization of someone with suicidal behavior. P17 uses this as an example of a label or recommendation a tool could give.}

\textbf{Some participants suggested abolishing the child welfare system and starting anew.} However, participants' thoughts on abolition were varied. At the end of one workshop, all four participants (all CPS workers) agreed that abolition would be the best solution (P29,P30,P31,P32). At the same time, a number of impacted parents who were very critical of the system said they did not think it should be abolished, but that it should be heavily reduced and reformed. Views on PRMs and abolition were also interestingly varied. P33 suggested that CPS should be reformed, but that PRMs should be abolished completely. P30 and P31 said to abolish CPS, but not PRMs: \textit{``I agree with tearing the system down. I just think that there's a place for the tools.''} P4 said that regardless of whether or not CPS is reformed or abolished, these are longer term changes and PRMs could help in the short term. See \cite{roberts2002shattered} or \cite{roberts2022torn} for more on child welfare abolition.

\section{Discussion}
\label{sec:discussion}

Here, we review novel suggestions and broader themes in Section~\ref{sec:results}, argue against the use of PRMs in child welfare, compare our study's approach with prior work, and highlight the suggestion to work in solidarity with impacted communities in the future.

\xhdr{Against predictive algorithms in CPS.} Our participants gave more novel suggestions and critical feedback than in prior participatory work with impacted communities and workers in CPS \cite{brown2019toward,saxena2020participatory,cheng2021soliciting,kawakami2022partnerships,cheng2022disparities,kawakami2022exploring}. For example, \citet{brown2019toward} suggest their participants' ``general distrust in the existing
system'' (which they somewhat vaguely describe as ``system-level concerns'') led to ``low comfort in algorithmic decision-making,'' and suggested these problems could be improved through ``greater transparency and improved communication strategies.'' Most of our participants also had ``low comfort'' in PRMs: They did not want them to be used. In Section~\ref{sec:participant-concerns}, our participants said PRMs would reify existing tendencies to punish instead of support poor, Black, and other marginalized families, and solidify existing power imbalances in CPS. Even if there are problems with PRMs, proponents argue for their use because they are better than any alternative, i.e. diagnostic checklists or nothing \cite{dare2016ethical}. In Section~\ref{sec:low-tech-alternatives}, however, participants gave low- and no-tech alternatives to address the problems motivating the use of PRMs: improved hiring, training, and working conditions; law and policy changes; giving money to families instead of CPS; and giving communities control of CPS. Overall, our participants thought PRMs are ``doing more harm than good'' and could be ``replaced by an equally viable low-tech or non-technological approach;'' thus, we argue that PRMs should not be used at all \cite{baumer2011implication}. 

\xhdr{Mitigating harms of PRMs.} Our participants also gave suggestions to mitigate the harms of PRMs (likely because they knew the above arguments are unlikely to stop agencies from using them). These suggestions largely differ from standard approaches to ``trustworthy'' AI. For example, participants did not ask for ``greater transparency'' around PRMs: They asked for regulations around how PRMs can and cannot be used, better evaluations of PRMs' impacts on communities, and more decisions about PRMs being made by impacted communities. Participants suggested that ``improved communication'' would not help either: although P14 suggested calling PRM labels \textit{``high need''} instead of \textit{``high risk''}, many participants said that it matters more who is giving the labels (CPS agencies) and what they are doing with them, e.g. surveillance instead of support.

\xhdr{Agreement between workers and parents.} Critiques of PRMs and CPS did not come only from parents, but from workers as well. This is surprising, because some described conflict between parents and workers. P32 said, \textit{``many of the white social workers have no knowledge of the suffering that goes on in the lives of the individuals they serve and cannot relate to their struggle.''} However, our worker and parent participants often agreed, and workers criticized CPS more than we expected. While this may be a result of self-selection bias, we believe it reveals a subset of CPS workers (not all of them) who work in CPS despite seeing how harmful it is to families (cf. \cite{copeland2021only}). These workers may be important accomplices for impacted communities organizing for change.

\xhdr{Why is it important to work with impacted stakeholders in child welfare?} For one, impacted stakeholders may generate ideas which researchers may not, due to lack of contextual knowledge or differing lived experiences. Many suggestions in Section~\ref{sec:new-uses} include these kinds of new design ideas. For another, impacted community perspectives are important in their own right, regardless of their value for novel research. Even when participants' suggestions are at odds academic work ---e.g. participants suggesting PRMs not use demographics, while prior work \cite{dwork2012fairness} suggests using demographics to mitigate disparities in PRMs,--- these suggestions are important because they reflect impacted stakeholders' perspectives. The general call to incorporate perspectives of impacted stakeholders into the design process \cite{zhu2018value,bjorgvinsson2010participatory} is heightened by the fact that the algorithms we focus on are used by governments which are accountable to the public \cite{lee2019webuildai,brown2019toward,holtenmoller2020shifting,saxena2020participatory,cheng2021soliciting}. If governments do not participate with impacted communities before they implement new technologies, they risk harming these communities, facing public scrutiny, or losing legitimacy \cite{pomeroy2019community,whittaker2018ai,propublica2016compas,lee2019webuildai}. Arnstein's Ladder of Civic Participation \cite{arnstein1969ladder} organizes participatory governance into levels of community involvement and empowerment. Lower levels involve consulting impacted communities on specific choices in later stages of development, but restricting communities' power to control whether public projects are implemented at all (which may verge on ``pseudo-participation'' \cite{palacin2020pseudoparticipation} or even ``participation-washing'' \cite{sloane2020participation}). Higher levels include empowering communities to negotiate the scope of public projects. Our work lies higher than prior work on Arnstein's Ladder \cite{arnstein1969ladder} in terms of scope, because we asked participants whether PRMs should be used in the first place, whereas prior work did not \cite{brown2019toward,saxena2020participatory,cheng2021soliciting,kawakami2022exploring,kawakami2022partnerships,cheng2022disparities}. However, prior work may have been limited in what kinds of choices they put ``on the table'' for stakeholders, because they worked with CPS agencies, which are either mandated to use, or have already chosen to use, algorithms \cite{saxena2021framework,delgado2021stakeholder}. Yet, by working with CPS agencies, prior work may have more influence over the design and use of algorithms (albeit in constrained ways). In our work, by contrast, we had more freedom to ask participants more basic questions about PRMs because we worked independently from a CPS agency. Yet, CPS agencies have no reason to listen to our suggestions. Thus, by Arnstein's measure \cite{arnstein1969ladder}, our work may not redistribute power to communities as much as prior work, because (by not working with a CPS agency) we do not have much power to change CPS policy on our own. This highlights not only tradeoffs in working with government agencies, but also the importance for researchers to collaborate with workers' and community groups who can apply power to influence agencies, while maintaining independence from agencies.

\xhdr{Work in solidarity with impacted communities.} Finally, our participants also suggested that researchers work in solidarity with impacted communities, even to oppose CPS agencies. This may have been overlooked in prior work because they centered public agencies. For example, \citet{brown2019toward} ask ``What can researchers and designers working in partnership with public service agencies do... to raise comfort levels among affected communities?'' then answer: ``Facilitate... positive relationships between child welfare workers and families.'' Yet, if researchers only encourage positive relationships, we may alienate people who have been harmed by CPS and do not want to stay positive. We should follow our participants' suggestion and work with impacted communities as ``academic accomplices'' \cite{asad2019academic}, whether that means evaluating CPS and workers, getting data in the hands of impacted communities (which is not always easy \cite{abdurahman2021calculating,sapien2016foil}), designing tools to recommend mandated reporters \textit{not} to report, joining with parents and advocates to fight against CPS agencies, or advocating for (non-technical) systemic changes. As groups like JMacForFamilies \cite{jmacforfamilies}, Movement for Family Power \cite{familypower}, the upEND Movement \cite{upEND}, and Rise \cite{rise} exemplify, impacted communities have been organizing themselves. Our participants suggest we work with them.

\begin{acks}
We thank Dr. Stevie Chancellor, Leah Ajmani, and our reviewers for their feedback, as well as Anushka Saxena and Janet Li for early work on this project. This work was supported by the National Science Foundation (NSF) under Award Nos. 2001851, 2000782, 1952085, the NSF Program on Fairness in AI in collaboration with Amazon under Award No.1939606, and Carnegie Mellon University Block Center for Technology and Society Award No. 53680.1.5007718.
\end{acks}

\newpage
\bibliographystyle{ACM-Reference-Format}
\bibliography{reference}


\begin{thebibliography}{138}

%%% the Web address.

\ifx \showCODEN    \undefined \def \showCODEN     #1{\unskip}     \fi
\ifx \showDOI      \undefined \def \showDOI       #1{#1}\fi
\ifx \showISBNx    \undefined \def \showISBNx     #1{\unskip}     \fi
\ifx \showISBNxiii \undefined \def \showISBNxiii  #1{\unskip}     \fi
\ifx \showISSN     \undefined \def \showISSN      #1{\unskip}     \fi
\ifx \showLCCN     \undefined \def \showLCCN      #1{\unskip}     \fi
\ifx \shownote     \undefined \def \shownote      #1{#1}          \fi
\ifx \showarticletitle \undefined \def \showarticletitle #1{#1}   \fi
\ifx \showURL      \undefined \def \showURL       {\relax}        \fi
% invisible to TeX
\providecommand\bibfield[2]{#2}
\providecommand\bibinfo[2]{#2}
\providecommand\natexlab[1]{#1}
\providecommand\showeprint[2][]{arXiv:#2}

\bibitem[\protect\citeauthoryear{Abdurahman}{Abdurahman}{2021a}]%
        {abdurahman2021body}
\bibfield{author}{\bibinfo{person}{J.~Khadijah Abdurahman}.}
  \bibinfo{year}{2021}\natexlab{a}.
\newblock \showarticletitle{A Body of Work That Cannot Be Ignored}.
\newblock \bibinfo{journal}{\emph{Logic Magazine}}  \bibinfo{volume}{15}
  (\bibinfo{year}{2021}).
\newblock
\urldef\tempurl%
\url{https://logicmag.io/beacons/a-body-of-work-that-cannot-be-ignored/}
\showURL{%
\tempurl}


\bibitem[\protect\citeauthoryear{Abdurahman}{Abdurahman}{2021b}]%
        {abdurahman2021calculating}
\bibfield{author}{\bibinfo{person}{J.~Khadijah Abdurahman}.}
  \bibinfo{year}{2021}\natexlab{b}.
\newblock \showarticletitle{Calculating the Souls of Black Folk: Predictive
  Analytics in the New York City Administration for Children’s Services}.
\newblock \bibinfo{journal}{\emph{Columbia Journal of Race and Law Forum}}
  \bibinfo{volume}{11}, \bibinfo{number}{4} (\bibinfo{year}{2021}),
  \bibinfo{pages}{75--110}.
\newblock
\urldef\tempurl%
\url{https://journals.library.columbia.edu/index.php/cjrl/article/view/8741}
\showURL{%
\tempurl}


\bibitem[\protect\citeauthoryear{Abdurahman, Lewis, McMillan, Montauban, Singh,
  and Newton}{Abdurahman et~al\mbox{.}}{2020}]%
        {webeimagining2020mothers}
\bibfield{author}{\bibinfo{person}{J.~Khadijah Abdurahman},
  \bibinfo{person}{Drew Lewis}, \bibinfo{person}{Joyce McMillan},
  \bibinfo{person}{Angeline Montauban}, \bibinfo{person}{Raquel Singh}, {and}
  \bibinfo{person}{Hope Newton}.} \bibinfo{year}{2020}\natexlab{}.
\newblock \bibinfo{title}{Minisode 4 - Mother's Day In The Trenches: Abolishing
  the Child Welfare System}.
\newblock
\newblock
\urldef\tempurl%
\url{https://americanassembly.org/wbi-podcast/minisode-child-welfare-ae7rh-84pj52-254c6-kxlej}
\showURL{%
\tempurl}


\bibitem[\protect\citeauthoryear{(ACLU)}{(ACLU)}{2021}]%
        {ccops}
\bibfield{author}{\bibinfo{person}{American Civil Liberties~Union (ACLU)}.}
  \bibinfo{year}{2021}\natexlab{}.
\newblock \bibinfo{title}{Community Control Over Police Surveillance (CCOPS)
  Model Bill}.
\newblock
  \bibinfo{howpublished}{\url{https://www.aclu.org/legal-document/community-control-over-police-surveillance-ccops-model-bill}}.
\newblock
\newblock
\shownote{Last Accessed: May 10, 2022.}


\bibitem[\protect\citeauthoryear{Adler}{Adler}{2001}]%
        {adler2001asfa}
\bibfield{author}{\bibinfo{person}{Libby Adler}.}
  \bibinfo{year}{2001}\natexlab{}.
\newblock \showarticletitle{The meanings of permanence: A critical analysis of
  the Adoption and Safe Families Act of 1997}.
\newblock \bibinfo{journal}{\emph{Harvard Journal on Legislation}}
  \bibinfo{volume}{38}, \bibinfo{number}{1} (\bibinfo{year}{2001}),
  \bibinfo{pages}{1---36}.
\newblock
\urldef\tempurl%
\url{https://ssrn.com/abstract=2045653}
\showURL{%
\tempurl}


\bibitem[\protect\citeauthoryear{Albright}{Albright}{2019}]%
        {albright2019judge}
\bibfield{author}{\bibinfo{person}{Alex Albright}.}
  \bibinfo{year}{2019}\natexlab{}.
\newblock \showarticletitle{If You Give a Judge a Risk Score: Evidence from
  Kentucky Bail Decisions}.
\newblock \bibinfo{journal}{\emph{Law, Economics, and Business Fellows’
  Discussion Paper Series 85}} (\bibinfo{year}{2019}).
\newblock


\bibitem[\protect\citeauthoryear{Allegheny County Department~of
  Human~Services}{Allegheny County Department~of Human~Services}{[n.d.]}]%
        {DHSResponse}
\bibfield{author}{\bibinfo{person}{(DHS) Allegheny County Department~of
  Human~Services}.} \bibinfo{year}{[n.d.]}\natexlab{}.
\newblock \bibinfo{booktitle}{\emph{DHS response to Automated Inequality by
  Virginia Eubanks}}.
\newblock
\urldef\tempurl%
\url{https://www.alleghenycounty.us/WorkArea/linkit.aspx?LinkIdentifier=id&ItemID=6442461672}
\showURL{%
\tempurl}


\bibitem[\protect\citeauthoryear{Antle, Barbee, Christensen, and
  Sullivan}{Antle et~al\mbox{.}}{2009}]%
        {antle2009prevention}
\bibfield{author}{\bibinfo{person}{B.~F. Antle}, \bibinfo{person}{A.~P.
  Barbee}, \bibinfo{person}{D.~N. Christensen}, {and} \bibinfo{person}{D.~J.
  Sullivan}.} \bibinfo{year}{2009}\natexlab{}.
\newblock \showarticletitle{The prevention of child maltreatment recidivism
  through the Solution-Based Casework model of child welfare practice}.
\newblock \bibinfo{journal}{\emph{Children and Youth Services Review}}
  \bibinfo{volume}{31}, \bibinfo{number}{12} (\bibinfo{year}{2009}),
  \bibinfo{pages}{1346–--1351}.
\newblock
\urldef\tempurl%
\url{https://doi.org/10.1016/j.childyouth.2009.06.008}
\showURL{%
\tempurl}


\bibitem[\protect\citeauthoryear{Arnstein}{Arnstein}{1969}]%
        {arnstein1969ladder}
\bibfield{author}{\bibinfo{person}{Sherry Arnstein}.}
  \bibinfo{year}{1969}\natexlab{}.
\newblock \showarticletitle{A Ladder of Citizen Participation}.
\newblock \bibinfo{journal}{\emph{Journal of the American Institute of
  Planners}} \bibinfo{number}{4} (\bibinfo{year}{1969}),
  \bibinfo{pages}{216---224}.
\newblock


\bibitem[\protect\citeauthoryear{Asad, Dombrowski, Costanza-Chock, Erete, and
  Harrington}{Asad et~al\mbox{.}}{2019}]%
        {asad2019academic}
\bibfield{author}{\bibinfo{person}{Mariam Asad}, \bibinfo{person}{Lynn
  Dombrowski}, \bibinfo{person}{Sasha Costanza-Chock}, \bibinfo{person}{Sheena
  Erete}, {and} \bibinfo{person}{Christina Harrington}.}
  \bibinfo{year}{2019}\natexlab{}.
\newblock \showarticletitle{Academic Accomplices: Practical Strategies for
  Research Justice}. In \bibinfo{booktitle}{\emph{Companion Publication of the
  2019 on Designing Interactive Systems Conference 2019 Companion}} (San Diego,
  CA, USA) \emph{(\bibinfo{series}{DIS '19 Companion})}.
  \bibinfo{publisher}{Association for Computing Machinery},
  \bibinfo{address}{New York, NY, USA}, \bibinfo{pages}{353–356}.
\newblock
\showISBNx{9781450362702}
\urldef\tempurl%
\url{https://doi.org/10.1145/3301019.3320001}
\showURL{%
\tempurl}


\bibitem[\protect\citeauthoryear{ASFA}{ASFA}{[n.d.]}]%
        {repealasfa}
\bibfield{author}{\bibinfo{person}{Repeal ASFA}.}
  \bibinfo{year}{[n.d.]}\natexlab{}.
\newblock \bibinfo{title}{Repeal ASFA}.
\newblock \bibinfo{howpublished}{\url{www.repealasfa.org}}.
\newblock
\newblock
\shownote{Last Accessed: May 10, 2022.}


\bibitem[\protect\citeauthoryear{Awad, Dsouza, Kim, Henrich, Shariff, Bonnefon,
  and Rahwan}{Awad et~al\mbox{.}}{2018}]%
        {awad2018moral}
\bibfield{author}{\bibinfo{person}{Edmond Awad}, \bibinfo{person}{Sohan
  Dsouza}, \bibinfo{person}{Jonathan Kim, Richard~Schulz},
  \bibinfo{person}{Joseph Henrich}, \bibinfo{person}{Azim Shariff},
  \bibinfo{person}{Jean-François Bonnefon}, {and} \bibinfo{person}{Iyad
  Rahwan}.} \bibinfo{year}{2018}\natexlab{}.
\newblock \showarticletitle{The Moral Machine experiment}.
\newblock \bibinfo{journal}{\emph{Nature}}  \bibinfo{volume}{563}
  (\bibinfo{year}{2018}), \bibinfo{pages}{59---64}.
\newblock
\urldef\tempurl%
\url{https://doi.org/10.1038/s41586-018-0637-6}
\showURL{%
\tempurl}


\bibitem[\protect\citeauthoryear{Baumer and Silberman}{Baumer and
  Silberman}{2011}]%
        {baumer2011implication}
\bibfield{author}{\bibinfo{person}{Eric~P.S. Baumer} {and}
  \bibinfo{person}{M.~Six Silberman}.} \bibinfo{year}{2011}\natexlab{}.
\newblock \showarticletitle{When the implication is not to design
  (technology)}. In \bibinfo{booktitle}{\emph{Proceedings of the 2011 SIGCHI
  conference on human factors in computing systems}}.
  \bibinfo{pages}{2271--2274}.
\newblock


\bibitem[\protect\citeauthoryear{Bechavod, Jung, and Wu}{Bechavod
  et~al\mbox{.}}{2020}]%
        {bechavod2020metric}
\bibfield{author}{\bibinfo{person}{Yahav Bechavod},
  \bibinfo{person}{Christopher Jung}, {and} \bibinfo{person}{Zhiwei~Steven
  Wu}.} \bibinfo{year}{2020}\natexlab{}.
\newblock \showarticletitle{Metric-free individual fairness in online
  learning}. In \bibinfo{booktitle}{\emph{Advances in Neural Information
  Processing Systems (NeurIPs)}}.
\newblock


\bibitem[\protect\citeauthoryear{Birhane}{Birhane}{2021}]%
        {birhane2021impossibility}
\bibfield{author}{\bibinfo{person}{Abeba Birhane}.}
  \bibinfo{year}{2021}\natexlab{}.
\newblock \showarticletitle{The Impossibility of Automating Ambiguity}.
\newblock \bibinfo{journal}{\emph{Artificial Life}} \bibinfo{volume}{27},
  \bibinfo{number}{1} (\bibinfo{year}{2021}), \bibinfo{pages}{44--61}.
\newblock
\urldef\tempurl%
\url{https://doi.org/10.1162/artl\_a\_00336}
\showURL{%
\tempurl}


\bibitem[\protect\citeauthoryear{Bjerknes, Kyng, Bjerknes, Ehn, Kyng, and
  Nygaard}{Bjerknes et~al\mbox{.}}{1987}]%
        {bjerknes1987computers}
\bibfield{author}{\bibinfo{person}{Ehn Bjerknes}, \bibinfo{person}{Nygaard
  Kyng}, \bibinfo{person}{Gro Bjerknes}, \bibinfo{person}{Pelle Ehn},
  \bibinfo{person}{Morten Kyng}, {and} \bibinfo{person}{Kristen Nygaard}.}
  \bibinfo{year}{1987}\natexlab{}.
\newblock \showarticletitle{Computers and democracy : a Scandinavian
  challenge}. \bibinfo{publisher}{Avebury}, \bibinfo{address}{Aldershot [Hants,
  England] ; Brookfield [Vt.], USA}.
\newblock
\showISBNx{0566054760}


\bibitem[\protect\citeauthoryear{Bj\"{o}rgvinsson, Ehn, and
  Hillgren}{Bj\"{o}rgvinsson et~al\mbox{.}}{2010}]%
        {bjorgvinsson2010participatory}
\bibfield{author}{\bibinfo{person}{Erling Bj\"{o}rgvinsson},
  \bibinfo{person}{Pelle Ehn}, {and} \bibinfo{person}{Per-Anders Hillgren}.}
  \bibinfo{year}{2010}\natexlab{}.
\newblock \showarticletitle{Participatory Design and ``Democratizing
  Innovation''}. In \bibinfo{booktitle}{\emph{Proceedings of the 11th Biennial
  Participatory Design Conference}} \emph{(\bibinfo{series}{PDC '10})}.
  \bibinfo{publisher}{Association for Computing Machinery},
  \bibinfo{address}{New York, NY, USA}, \bibinfo{pages}{41–50}.
\newblock
\showISBNx{9781450301312}
\urldef\tempurl%
\url{https://doi.org/10.1145/1900441.1900448}
\showDOI{\tempurl}


\bibitem[\protect\citeauthoryear{Braun and Clarke}{Braun and Clarke}{2006}]%
        {braun2006thematic}
\bibfield{author}{\bibinfo{person}{Virginia Braun} {and}
  \bibinfo{person}{Victoria Clarke}.} \bibinfo{year}{2006}\natexlab{}.
\newblock \showarticletitle{Using thematic analysis in psychology}.
\newblock \bibinfo{journal}{\emph{Qualitative Research in Psychology}}
  \bibinfo{volume}{3}, \bibinfo{number}{2} (\bibinfo{year}{2006}),
  \bibinfo{pages}{77--101}.
\newblock
\urldef\tempurl%
\url{https://doi.org/10.1191/1478088706qp063oa}
\showDOI{\tempurl}
\showeprint{https://www.tandfonline.com/doi/pdf/10.1191/1478088706qp063oa}


\bibitem[\protect\citeauthoryear{Brayne}{Brayne}{2017}]%
        {brayne2017policing}
\bibfield{author}{\bibinfo{person}{Sarah Brayne}.}
  \bibinfo{year}{2017}\natexlab{}.
\newblock \showarticletitle{Big Data Surveillance: The Case of Policing}.
\newblock \bibinfo{journal}{\emph{American Sociological Review}}
  \bibinfo{volume}{82}, \bibinfo{number}{5} (\bibinfo{year}{2017}),
  \bibinfo{pages}{977--1008}.
\newblock
\urldef\tempurl%
\url{https://doi.org/10.1177/0003122417725865}
\showDOI{\tempurl}


\bibitem[\protect\citeauthoryear{Brown, Chouldechova, Putnam{-}Hornstein,
  Tobin, and Vaithianathan}{Brown et~al\mbox{.}}{2019}]%
        {brown2019toward}
\bibfield{author}{\bibinfo{person}{Anna Brown}, \bibinfo{person}{Alexandra
  Chouldechova}, \bibinfo{person}{Emily Putnam{-}Hornstein},
  \bibinfo{person}{Andrew Tobin}, {and} \bibinfo{person}{Rhema Vaithianathan}.}
  \bibinfo{year}{2019}\natexlab{}.
\newblock \showarticletitle{Toward Algorithmic Accountability in Public
  Services: A Qualitative Study of Affected Community Perspectives on
  Algorithmic Decision-making in Child Welfare Services}. In
  \bibinfo{booktitle}{\emph{Proceedings of the 2019 {CHI} Conference on Human
  Factors in Computing Systems, {CHI} 2019, Glasgow, Scotland, UK, May 04-09,
  2019}}. \bibinfo{pages}{41}.
\newblock
\urldef\tempurl%
\url{https://doi.org/10.1145/3290605.3300271}
\showDOI{\tempurl}


\bibitem[\protect\citeauthoryear{Burton and Montauban}{Burton and
  Montauban}{2021}]%
        {burton2021toward}
\bibfield{author}{\bibinfo{person}{Angela~Olivia Burton} {and}
  \bibinfo{person}{Angeline Montauban}.} \bibinfo{year}{2021}\natexlab{}.
\newblock \showarticletitle{Toward Community Control of Child Welfare Funding:
  Repeal the Child Abuse Prevention and Treatment Act and Delink Child
  Protection from Family Well-being}.
\newblock \bibinfo{journal}{\emph{Columbia Journal of Race and Law Forum}}
  \bibinfo{volume}{11}, \bibinfo{number}{3} (\bibinfo{year}{2021}).
\newblock
\urldef\tempurl%
\url{https://doi.org/10.52214/cjrl.v11i3.8747}
\showURL{%
\tempurl}


\bibitem[\protect\citeauthoryear{Carton, Helsby, Joseph, Mahmud, Park, Walsh,
  Cody, Patterson, Haynes, and Ghani}{Carton et~al\mbox{.}}{2016}]%
        {carton2016identifying}
\bibfield{author}{\bibinfo{person}{Samuel Carton}, \bibinfo{person}{Jennifer
  Helsby}, \bibinfo{person}{Kenneth Joseph}, \bibinfo{person}{Ayesha Mahmud},
  \bibinfo{person}{Youngsoo Park}, \bibinfo{person}{Joe Walsh},
  \bibinfo{person}{Crystal Cody}, \bibinfo{person}{CPT~Estella Patterson},
  \bibinfo{person}{Lauren Haynes}, {and} \bibinfo{person}{Rayid Ghani}.}
  \bibinfo{year}{2016}\natexlab{}.
\newblock \showarticletitle{Identifying Police Officers at Risk of Adverse
  Events}. In \bibinfo{booktitle}{\emph{Proceedings of the 22nd ACM SIGKDD
  International Conference on Knowledge Discovery and Data Mining}} (San
  Francisco, California, USA) \emph{(\bibinfo{series}{KDD '16})}.
  \bibinfo{publisher}{Association for Computing Machinery},
  \bibinfo{address}{New York, NY, USA}, \bibinfo{pages}{67–76}.
\newblock
\showISBNx{9781450342322}
\urldef\tempurl%
\url{https://doi.org/10.1145/2939672.2939698}
\showDOI{\tempurl}


\bibitem[\protect\citeauthoryear{Cecilia, Shion, Marina, Michael, and
  Gina}{Cecilia et~al\mbox{.}}{2022}]%
        {aragon2022human}
\bibfield{author}{\bibinfo{person}{Aragon Cecilia}, \bibinfo{person}{Guha
  Shion}, \bibinfo{person}{Kogan Marina}, \bibinfo{person}{Muller Michael},
  {and} \bibinfo{person}{Neff Gina}.} \bibinfo{year}{2022}\natexlab{}.
\newblock \bibinfo{booktitle}{\emph{Human-Centered Data Science : An
  Introduction}}.
\newblock \bibinfo{publisher}{The MIT Press}.
\newblock
\showISBNx{9780262543217}


\bibitem[\protect\citeauthoryear{Chancellor}{Chancellor}{2022}]%
        {chancellor2022practices}
\bibfield{author}{\bibinfo{person}{Stevie Chancellor}.}
  \bibinfo{year}{2022}\natexlab{}.
\newblock \bibinfo{title}{Towards Practices for Human-Centered Machine
  Learning}.
\newblock
\newblock
\urldef\tempurl%
\url{https://arxiv.org/abs/2203.00432}
\showURL{%
\tempurl}


\bibitem[\protect\citeauthoryear{Chancellor, Baumer, and
  De~Choudhury}{Chancellor et~al\mbox{.}}{2019}]%
        {chancellor2019who}
\bibfield{author}{\bibinfo{person}{Stevie Chancellor}, \bibinfo{person}{Eric
  P.~S. Baumer}, {and} \bibinfo{person}{Munmun De~Choudhury}.}
  \bibinfo{year}{2019}\natexlab{}.
\newblock \showarticletitle{Who is the "Human" in Human-Centered Machine
  Learning: The Case of Predicting Mental Health from Social Media}.
\newblock \bibinfo{journal}{\emph{Proc. ACM Hum.-Comput. Interact.}}
  \bibinfo{volume}{3}, \bibinfo{number}{CSCW} (\bibinfo{year}{2019}),
  \bibinfo{numpages}{32}~pages.
\newblock
\urldef\tempurl%
\url{https://doi.org/10.1145/3359249}
\showDOI{\tempurl}


\bibitem[\protect\citeauthoryear{Cheng, Stapleton, Kawakami, Sivaraman, Cheng,
  Qing, Perer, Holstein, Wu, and Zhu}{Cheng et~al\mbox{.}}{2022}]%
        {cheng2022disparities}
\bibfield{author}{\bibinfo{person}{Hao-Fei Cheng}, \bibinfo{person}{Logan
  Stapleton}, \bibinfo{person}{Anna Kawakami}, \bibinfo{person}{Venkat
  Sivaraman}, \bibinfo{person}{Yanghuidi Cheng}, \bibinfo{person}{Diana Qing},
  \bibinfo{person}{Adam Perer}, \bibinfo{person}{Kenneth Holstein},
  \bibinfo{person}{Zhiwei~Steven Wu}, {and} \bibinfo{person}{Haiyi Zhu}.}
  \bibinfo{year}{2022}\natexlab{}.
\newblock \showarticletitle{How Child Welfare Workers Reduce Racial Disparities
  in Algorithmic Decisions}. In \bibinfo{booktitle}{\emph{Proceedings of the
  2022 CHI Conference on Human Factors in Computing Systems}}.
\newblock


\bibitem[\protect\citeauthoryear{Cheng, Stapleton, Wang, Bullock, Chouldechova,
  Wu, and Zhu}{Cheng et~al\mbox{.}}{2021}]%
        {cheng2021soliciting}
\bibfield{author}{\bibinfo{person}{Hao-Fei Cheng}, \bibinfo{person}{Logan
  Stapleton}, \bibinfo{person}{Ruiqi Wang}, \bibinfo{person}{Paige Bullock},
  \bibinfo{person}{Alexandra Chouldechova}, \bibinfo{person}{Zhiwei
  Steven~Steven Wu}, {and} \bibinfo{person}{Haiyi Zhu}.}
  \bibinfo{year}{2021}\natexlab{}.
\newblock \showarticletitle{Soliciting Stakeholders’ Fairness Notions in
  Child Maltreatment Predictive Systems}. In
  \bibinfo{booktitle}{\emph{Proceedings of the 2021 CHI Conference on Human
  Factors in Computing Systems}}. \bibinfo{publisher}{Association for Computing
  Machinery}, \bibinfo{address}{New York, NY, USA}, Article
  \bibinfo{articleno}{390}, \bibinfo{numpages}{17}~pages.
\newblock
\showISBNx{9781450380966}
\urldef\tempurl%
\url{https://doi.org/10.1145/3411764.3445308}
\showURL{%
\tempurl}


\bibitem[\protect\citeauthoryear{Chouldechova, Benavides-Prado, Fialko, and
  Vaithianathan}{Chouldechova et~al\mbox{.}}{2018}]%
        {chouldechova2018case}
\bibfield{author}{\bibinfo{person}{Alexandra Chouldechova},
  \bibinfo{person}{Diana Benavides-Prado}, \bibinfo{person}{Oleksandr Fialko},
  {and} \bibinfo{person}{Rhema Vaithianathan}.}
  \bibinfo{year}{2018}\natexlab{}.
\newblock \showarticletitle{A case study of algorithm-assisted decision making
  in child maltreatment hotline screening decisions}. In
  \bibinfo{booktitle}{\emph{Conference on Fairness, Accountability and
  Transparency}}. PMLR, \bibinfo{pages}{134--148}.
\newblock


\bibitem[\protect\citeauthoryear{Church and Fairchild}{Church and
  Fairchild}{2017}]%
        {church2017silver}
\bibfield{author}{\bibinfo{person}{Christopher~E. Church} {and}
  \bibinfo{person}{Amanda~J. Fairchild}.} \bibinfo{year}{2017}\natexlab{}.
\newblock \showarticletitle{In Search of a Silver Bullet: Child Welfare's
  Embrace of Predictive Analytics}.
\newblock \bibinfo{journal}{\emph{Juvenile and Family Court Journal}}
  \bibinfo{volume}{68}, \bibinfo{number}{1} (\bibinfo{year}{2017}),
  \bibinfo{pages}{67--81}.
\newblock
\urldef\tempurl%
\url{https://onlinelibrary.wiley.com/doi/abs/10.1111/jfcj.12086}
\showURL{%
\tempurl}


\bibitem[\protect\citeauthoryear{Collins}{Collins}{1997}]%
        {collins1997standpoint}
\bibfield{author}{\bibinfo{person}{Patricia~Hill Collins}.}
  \bibinfo{year}{1997}\natexlab{}.
\newblock \showarticletitle{Comment on Hekman's "Truth and Method: Feminist
  Standpoint Theory Revisited": Where's the Power?}
\newblock \bibinfo{journal}{\emph{Signs}} \bibinfo{volume}{22},
  \bibinfo{number}{2} (\bibinfo{year}{1997}), \bibinfo{pages}{375--381}.
\newblock
\urldef\tempurl%
\url{http://www.jstor.org/stable/3175278}
\showURL{%
\tempurl}


\bibitem[\protect\citeauthoryear{Connects}{Connects}{[n.d.]}]%
        {eckerdrsf}
\bibfield{author}{\bibinfo{person}{Eckerd Connects}.}
  \bibinfo{year}{[n.d.]}\natexlab{}.
\newblock \bibinfo{title}{ECKERD RAPID SAFETY FEEDBACK}.
\newblock
  \bibinfo{howpublished}{\url{https://eckerd.org/family-children-services/ersf/}}.
\newblock
\newblock
\shownote{Online; accessed 8-September-2021.}


\bibitem[\protect\citeauthoryear{Copeland}{Copeland}{2021}]%
        {copeland2021only}
\bibfield{author}{\bibinfo{person}{Victoria Copeland}.}
  \bibinfo{year}{2021}\natexlab{}.
\newblock \showarticletitle{``It's the Only System We've Got'': Exploring
  Emergency Response Decision-Making in Child Welfare}.
\newblock \bibinfo{journal}{\emph{Columbia Journal of Race and Law Forum}}
  \bibinfo{volume}{11} (\bibinfo{year}{2021}).
\newblock
\urldef\tempurl%
\url{https://doi.org/10.52214/cjrl.v11i3.8740}
\showURL{%
\tempurl}


\bibitem[\protect\citeauthoryear{Coston, Mishler, Kennedy, and
  Chouldechova}{Coston et~al\mbox{.}}{2020}]%
        {coston2020counterfactual}
\bibfield{author}{\bibinfo{person}{Amanda Coston}, \bibinfo{person}{Alan
  Mishler}, \bibinfo{person}{Edward~H Kennedy}, {and}
  \bibinfo{person}{Alexandra Chouldechova}.} \bibinfo{year}{2020}\natexlab{}.
\newblock \showarticletitle{Counterfactual risk assessments, evaluation, and
  fairness}. In \bibinfo{booktitle}{\emph{Proceedings of the 2020 Conference on
  Fairness, Accountability, and Transparency}}. \bibinfo{pages}{582--593}.
\newblock


\bibitem[\protect\citeauthoryear{Dare and Gambrill}{Dare and Gambrill}{2016}]%
        {dare2016ethical}
\bibfield{author}{\bibinfo{person}{Tim Dare} {and} \bibinfo{person}{Eileen
  Gambrill}.} \bibinfo{year}{2016}\natexlab{}.
\newblock \showarticletitle{Ethical Analysis: Predictive Risk Models at Call
  Screening for Allegheny County}.
\newblock  (\bibinfo{year}{2016}).
\newblock
\urldef\tempurl%
\url{https://www.alleghenycountyanalytics.us/wp-content/uploads/2019/05/Ethical-Analysis-16-ACDHS-26_PredictiveRisk_Package_050119_FINAL-2.pdf}
\showURL{%
\tempurl}
\newblock
\shownote{Online; accessed 8-September-2021.}


\bibitem[\protect\citeauthoryear{De-Arteaga, Fogliato, and
  Chouldechova}{De-Arteaga et~al\mbox{.}}{2020}]%
        {De-Arteaga2020case}
\bibfield{author}{\bibinfo{person}{Maria De-Arteaga}, \bibinfo{person}{Riccardo
  Fogliato}, {and} \bibinfo{person}{Alexandra Chouldechova}.}
  \bibinfo{year}{2020}\natexlab{}.
\newblock \showarticletitle{A case for humans-in-the-loop: Decisions in the
  presence of erroneous algorithmic scores}. In
  \bibinfo{booktitle}{\emph{Proceedings of the 2020 CHI Conference on Human
  Factors in Computing Systems}}. \bibinfo{pages}{1--12}.
\newblock
\urldef\tempurl%
\url{https://papers.ssrn.com/sol3/papers.cfm?abstract_id=4050125}
\showURL{%
\tempurl}


\bibitem[\protect\citeauthoryear{Delgado, Yang, Madaio, and Yang}{Delgado
  et~al\mbox{.}}{2021}]%
        {delgado2021stakeholder}
\bibfield{author}{\bibinfo{person}{Fernando Delgado}, \bibinfo{person}{Stephen
  Yang}, \bibinfo{person}{Michael Madaio}, {and} \bibinfo{person}{Qian Yang}.}
  \bibinfo{year}{2021}\natexlab{}.
\newblock \showarticletitle{Stakeholder Participation in AI: Beyond ``Add
  Diverse Stakeholders and Stir''}.
\newblock \bibinfo{journal}{\emph{arXiv preprint}}
  \bibinfo{volume}{arXiv:2111.01122} (\bibinfo{year}{2021}).
\newblock


\bibitem[\protect\citeauthoryear{Dettlaff and Boyd}{Dettlaff and Boyd}{2020}]%
        {dettlaff2020racial}
\bibfield{author}{\bibinfo{person}{Alan~J Dettlaff} {and}
  \bibinfo{person}{Reiko Boyd}.} \bibinfo{year}{2020}\natexlab{}.
\newblock \showarticletitle{Racial disproportionality and disparities in the
  child welfare system: Why do they exist, and what can be done to address
  them?}
\newblock \bibinfo{journal}{\emph{The ANNALS of the American Academy of
  Political and Social Science}} \bibinfo{volume}{692}, \bibinfo{number}{1}
  (\bibinfo{year}{2020}), \bibinfo{pages}{253--274}.
\newblock


\bibitem[\protect\citeauthoryear{Dettlaff, Rivaux, Baumann, Fluke, Rycraft, and
  James}{Dettlaff et~al\mbox{.}}{2011}]%
        {dettlaff2011disentangling}
\bibfield{author}{\bibinfo{person}{Alan~J Dettlaff},
  \bibinfo{person}{Stephanie~L Rivaux}, \bibinfo{person}{Donald~J Baumann},
  \bibinfo{person}{John~D Fluke}, \bibinfo{person}{Joan~R Rycraft}, {and}
  \bibinfo{person}{Joyce James}.} \bibinfo{year}{2011}\natexlab{}.
\newblock \showarticletitle{Disentangling substantiation: The influence of
  race, income, and risk on the substantiation decision in child welfare}.
\newblock \bibinfo{journal}{\emph{Children and Youth Services Review}}
  \bibinfo{volume}{33}, \bibinfo{number}{9} (\bibinfo{year}{2011}),
  \bibinfo{pages}{1630--1637}.
\newblock


\bibitem[\protect\citeauthoryear{Drake and Jonson-Reid}{Drake and
  Jonson-Reid}{2014}]%
        {drake2014poverty}
\bibfield{author}{\bibinfo{person}{Brett Drake} {and} \bibinfo{person}{Melissa
  Jonson-Reid}.} \bibinfo{year}{2014}\natexlab{}.
\newblock \bibinfo{booktitle}{\emph{Poverty and Child Maltreatment}}.
\newblock \bibinfo{publisher}{Springer Netherlands}, \bibinfo{pages}{131--148}.
\newblock
\showISBNx{978-94-007-7208-3}
\urldef\tempurl%
\url{https://doi.org/10.1007/978-94-007-7208-3_7}
\showDOI{\tempurl}


\bibitem[\protect\citeauthoryear{Drake, Jonson-Reid, Gandarilla~Ocampo,
  Morrison, and Dvalishvili}{Drake et~al\mbox{.}}{2020}]%
        {drake2020practical}
\bibfield{author}{\bibinfo{person}{Brett Drake}, \bibinfo{person}{Melissa
  Jonson-Reid}, \bibinfo{person}{Mar\'{i}a Gandarilla~Ocampo},
  \bibinfo{person}{Maria Morrison}, {and} \bibinfo{person}{Darejan~(Daji)
  Dvalishvili}.} \bibinfo{year}{2020}\natexlab{}.
\newblock \showarticletitle{A Practical Framework for Considering the Use of
  Predictive Risk Modeling in Child Welfare}.
\newblock \bibinfo{journal}{\emph{Annals of the American Academy of Political
  Science \& Social Science (AAPSS)}}  \bibinfo{volume}{692}
  (\bibinfo{year}{2020}).
\newblock


\bibitem[\protect\citeauthoryear{Dwork, Hardt, Pitassi, Reingold, and
  Zemel}{Dwork et~al\mbox{.}}{2012}]%
        {dwork2012fairness}
\bibfield{author}{\bibinfo{person}{Cynthia Dwork}, \bibinfo{person}{Moritz
  Hardt}, \bibinfo{person}{Toniann Pitassi}, \bibinfo{person}{Omer Reingold},
  {and} \bibinfo{person}{Richard Zemel}.} \bibinfo{year}{2012}\natexlab{}.
\newblock \showarticletitle{Fairness through awareness}. In
  \bibinfo{booktitle}{\emph{Proceedings of the 3rd innovations in theoretical
  computer science conference}}. \bibinfo{pages}{214--226}.
\newblock


\bibitem[\protect\citeauthoryear{Edwards and Wildeman}{Edwards and
  Wildeman}{2020}]%
        {edwards2020characteristics}
\bibfield{author}{\bibinfo{person}{Frank Edwards} {and}
  \bibinfo{person}{Christopher Wildeman}.} \bibinfo{year}{2020}\natexlab{}.
\newblock \bibinfo{title}{Characteristics of the Front-Line Child Welfare
  Workforce}.
\newblock
\newblock
\urldef\tempurl%
\url{https://www.socialserviceworkforce.org/resources/characteristics-front-line-child-welfare-workforce}
\showURL{%
\tempurl}


\bibitem[\protect\citeauthoryear{Ehn}{Ehn}{1993}]%
        {ehn1993scandinavian}
\bibfield{author}{\bibinfo{person}{Paul Ehn}.} \bibinfo{year}{1993}\natexlab{}.
\newblock \bibinfo{booktitle}{\emph{Scandinavian Design: On Participation and
  Skill}}.
\newblock \bibinfo{publisher}{Oxford University Press}.
\newblock
\urldef\tempurl%
\url{https://oxford.universitypressscholarship.com/view/10.1093/oso/9780195075106.001.0001/isbn-9780195075106-book-part-8}
\showURL{%
\tempurl}


\bibitem[\protect\citeauthoryear{Elish}{Elish}{2019}]%
        {elish2019crumple}
\bibfield{author}{\bibinfo{person}{Madeleine~Claire Elish}.}
  \bibinfo{year}{2019}\natexlab{}.
\newblock \showarticletitle{Moral crumple zones: Cautionary tales in
  human-robot interaction}.
\newblock \bibinfo{journal}{\emph{Science, Technology, and Society}}
  \bibinfo{volume}{5} (\bibinfo{year}{2019}), \bibinfo{pages}{40--60}.
\newblock


\bibitem[\protect\citeauthoryear{Eubanks}{Eubanks}{2018}]%
        {eubanks2018automating}
\bibfield{author}{\bibinfo{person}{Virginia Eubanks}.}
  \bibinfo{year}{2018}\natexlab{}.
\newblock \bibinfo{booktitle}{\emph{Automating Inequality: How High-tech Tools
  Profile, Police, and Punish the Poor}}.
\newblock \bibinfo{publisher}{New York, NY: St. Martin's Press}.
\newblock


\bibitem[\protect\citeauthoryear{Freire}{Freire}{1972}]%
        {freire1972pedagogy}
\bibfield{author}{\bibinfo{person}{Paulo Freire}.}
  \bibinfo{year}{1972}\natexlab{}.
\newblock \bibinfo{booktitle}{\emph{Pedagogy of the Oppressed}}.
\newblock \bibinfo{publisher}{Herder and Herder}.
\newblock


\bibitem[\protect\citeauthoryear{Glaberson}{Glaberson}{2019}]%
        {glaberson2019coding}
\bibfield{author}{\bibinfo{person}{Stephanie~K. Glaberson}.}
  \bibinfo{year}{2019}\natexlab{}.
\newblock \showarticletitle{Coding Over the Cracks: Predictive Analytics and
  Child Protection}.
\newblock \bibinfo{journal}{\emph{Fordham Urban Law Journal}}
  (\bibinfo{year}{2019}).
\newblock
\urldef\tempurl%
\url{https://ir.lawnet.fordham.edu/ulj/vol46/iss2/3}
\showURL{%
\tempurl}


\bibitem[\protect\citeauthoryear{Goldhaber-Fiebert and
  Prince}{Goldhaber-Fiebert and Prince}{2019}]%
        {goldhaber2019impact}
\bibfield{author}{\bibinfo{person}{Jeremy~D Goldhaber-Fiebert} {and}
  \bibinfo{person}{Lea Prince}.} \bibinfo{year}{2019}\natexlab{}.
\newblock \showarticletitle{Impact evaluation of a predictive risk modeling
  tool for Allegheny county’s child welfare office}.
\newblock \bibinfo{journal}{\emph{Pittsburgh: Allegheny County}}
  (\bibinfo{year}{2019}).
\newblock


\bibitem[\protect\citeauthoryear{Green}{Green}{2021}]%
        {green2021flaws}
\bibfield{author}{\bibinfo{person}{Ben Green}.}
  \bibinfo{year}{2021}\natexlab{}.
\newblock \showarticletitle{The Flaws of Policies Requiring Human Oversight of
  Government Algorithms}.
\newblock \bibinfo{journal}{\emph{SSRN}} (\bibinfo{year}{2021}).
\newblock
\urldef\tempurl%
\url{http://dx.doi.org/10.2139/ssrn.3921216}
\showURL{%
\tempurl}


\bibitem[\protect\citeauthoryear{Gregory}{Gregory}{2003}]%
        {gregory2003scandinavian}
\bibfield{author}{\bibinfo{person}{Judith Gregory}.}
  \bibinfo{year}{2003}\natexlab{}.
\newblock \showarticletitle{Scandinavian Approaches to Participatory Design}.
\newblock \bibinfo{journal}{\emph{International Journal of Engineering
  Education}}  \bibinfo{volume}{19} (\bibinfo{year}{2003}).
\newblock


\bibitem[\protect\citeauthoryear{Grgi\'{c}-Hla\v{c}a, Zafar, Gummadi, and
  Weller}{Grgi\'{c}-Hla\v{c}a et~al\mbox{.}}{2018}]%
        {grgichlaca2018}
\bibfield{author}{\bibinfo{person}{Nina Grgi\'{c}-Hla\v{c}a},
  \bibinfo{person}{Muhammad~Bilal Zafar}, \bibinfo{person}{Krishna~P. Gummadi},
  {and} \bibinfo{person}{Adrian Weller}.} \bibinfo{year}{2018}\natexlab{}.
\newblock \showarticletitle{Beyond distributive fairness in algorithmic
  decision making: Feature selection for procedurally fair learning}. In
  \bibinfo{booktitle}{\emph{Proceedings of the 32nd AAAI Conference on
  Artificial Intelligence}}.
\newblock


\bibitem[\protect\citeauthoryear{Halfaker and Geiger}{Halfaker and
  Geiger}{2020}]%
        {halfaker2020ores}
\bibfield{author}{\bibinfo{person}{Aaron Halfaker} {and}
  \bibinfo{person}{R~Stuart Geiger}.} \bibinfo{year}{2020}\natexlab{}.
\newblock \showarticletitle{Ores: Lowering barriers with participatory machine
  learning in wikipedia}.
\newblock \bibinfo{journal}{\emph{Proceedings of the ACM on Human-Computer
  Interaction}} \bibinfo{volume}{4}, \bibinfo{number}{CSCW2}
  (\bibinfo{year}{2020}), \bibinfo{pages}{1--37}.
\newblock


\bibitem[\protect\citeauthoryear{Harding}{Harding}{2004}]%
        {harding2004feminist}
\bibfield{author}{\bibinfo{person}{S.G. Harding}.}
  \bibinfo{year}{2004}\natexlab{}.
\newblock \bibinfo{booktitle}{\emph{The Feminist Standpoint Theory Reader:
  Intellectual and Political Controversies}}.
\newblock \bibinfo{publisher}{Routledge}.
\newblock
\showISBNx{9780415945011}
\urldef\tempurl%
\url{https://books.google.com/books?id=qmSySHvIy5IC}
\showURL{%
\tempurl}


\bibitem[\protect\citeauthoryear{Harding}{Harding}{2016}]%
        {harding2016standpoint}
\bibfield{author}{\bibinfo{person}{Sandra Harding}.}
  \bibinfo{year}{2016}\natexlab{}.
\newblock \bibinfo{title}{Sandra Harding: On Standpoint Theory's History and
  Controversial Reception}.
\newblock
  \bibinfo{howpublished}{\url{https://www.youtube.com/watch?v=xOAMc12PqmI}}.
\newblock


\bibitem[\protect\citeauthoryear{Harrington, Erete, and Piper}{Harrington
  et~al\mbox{.}}{2019}]%
        {harrington2019deconstructing}
\bibfield{author}{\bibinfo{person}{Christina Harrington},
  \bibinfo{person}{Sheena Erete}, {and} \bibinfo{person}{Anne~Marie Piper}.}
  \bibinfo{year}{2019}\natexlab{}.
\newblock \showarticletitle{Deconstructing community-based collaborative
  design: Towards more equitable participatory design engagements}.
\newblock \bibinfo{journal}{\emph{Proceedings of the ACM on Human-Computer
  Interaction}} \bibinfo{volume}{3}, \bibinfo{number}{CSCW}
  (\bibinfo{year}{2019}), \bibinfo{pages}{1--25}.
\newblock


\bibitem[\protect\citeauthoryear{Holstein, McLaren, and Aleven}{Holstein
  et~al\mbox{.}}{2018}]%
        {holstein2018student}
\bibfield{author}{\bibinfo{person}{Kenneth Holstein}, \bibinfo{person}{Bruce~M
  McLaren}, {and} \bibinfo{person}{Vincent Aleven}.}
  \bibinfo{year}{2018}\natexlab{}.
\newblock \showarticletitle{Student learning benefits of a mixed-reality
  teacher awareness tool in AI-enhanced classrooms}. In
  \bibinfo{booktitle}{\emph{International Conference on Artificial Intelligence
  in Education}}. Springer, \bibinfo{pages}{154--168}.
\newblock


\bibitem[\protect\citeauthoryear{{Holten M{\o}ller}, Shklovski, and
  Hildebrandt}{{Holten M{\o}ller} et~al\mbox{.}}{2020}]%
        {holtenmoller2020shifting}
\bibfield{author}{\bibinfo{person}{Naja {Holten M{\o}ller}},
  \bibinfo{person}{Irina Shklovski}, {and} \bibinfo{person}{Thomas~T.
  Hildebrandt}.} \bibinfo{year}{2020}\natexlab{}.
\newblock \showarticletitle{{Shifting concepts of value: Designing algorithmic
  decision-support systems for public services}}.
\newblock \bibinfo{journal}{\emph{NordiCHI}} (\bibinfo{year}{2020}),
  \bibinfo{pages}{1--12}.
\newblock
\showISBNx{9781450375795}
\urldef\tempurl%
\url{https://doi.org/10.1145/3419249.3420149}
\showDOI{\tempurl}


\bibitem[\protect\citeauthoryear{Hurley}{Hurley}{2018}]%
        {hurley2018algorithm}
\bibfield{author}{\bibinfo{person}{Dan Hurley}.}
  \bibinfo{year}{2018}\natexlab{}.
\newblock \showarticletitle{Can an Algorithm Tell When Kids Are in Danger?}
\newblock \bibinfo{journal}{\emph{New York Times Magazine}}
  (\bibinfo{year}{2018}).
\newblock
\urldef\tempurl%
\url{nytimes.com/2018/01/02/magazine/can-an-algorithm-tell-when-kids-are-in-danger.html}
\showURL{%
\tempurl}
\newblock
\shownote{Online; accessed 8-September-2021.}


\bibitem[\protect\citeauthoryear{Ilvento}{Ilvento}{2019}]%
        {ilvento2019metric}
\bibfield{author}{\bibinfo{person}{Christina Ilvento}.}
  \bibinfo{year}{2019}\natexlab{}.
\newblock \showarticletitle{Metric Learning for Individual Fairness}.
\newblock \bibinfo{journal}{\emph{arXiv preprint}}
  \bibinfo{volume}{arXiv:1906.00250} (\bibinfo{year}{2019}).
\newblock


\bibitem[\protect\citeauthoryear{Institute}{Institute}{[n.d.]}]%
        {sas}
\bibfield{author}{\bibinfo{person}{SAS Institute}.}
  \bibinfo{year}{[n.d.]}\natexlab{}.
\newblock \bibinfo{title}{Analytics for Child Well-Being}.
\newblock
  \bibinfo{howpublished}{\url{https://www.sas.com/en_us/software/analytics-for-child-well-being.html}}.
\newblock
\newblock
\shownote{Online; accessed 12-December-2021.}


\bibitem[\protect\citeauthoryear{Jackson and Marx}{Jackson and Marx}{2017}]%
        {jackson2017illinois}
\bibfield{author}{\bibinfo{person}{David Jackson} {and} \bibinfo{person}{Gary
  Marx}.} \bibinfo{year}{2017}\natexlab{}.
\newblock \showarticletitle{Data mining program designed to predict child abuse
  proves unreliable, DCFS says}.
\newblock \bibinfo{journal}{\emph{Chicago Tribune}} (\bibinfo{year}{2017}).
\newblock
\urldef\tempurl%
\url{https://www.chicagotribune.com/investigations/ct-dcfs-eckerd-met-20171206-story.html}
\showURL{%
\tempurl}
\newblock
\shownote{Online; accessed 6-January-2022.}


\bibitem[\protect\citeauthoryear{JMacForFamilies}{JMacForFamilies}{[n.d.]}]%
        {jmacforfamilies}
\bibfield{author}{\bibinfo{person}{JMacForFamilies}.}
  \bibinfo{year}{[n.d.]}\natexlab{}.
\newblock \bibinfo{title}{JMacForFamilies}.
\newblock \bibinfo{howpublished}{\url{https://jmacforfamilies.org/plan}}.
\newblock
\newblock
\shownote{Last Accessed: May 10, 2022.}


\bibitem[\protect\citeauthoryear{Johnston, Blessenohl, and Vayanos}{Johnston
  et~al\mbox{.}}{2020}]%
        {johnston2020preference}
\bibfield{author}{\bibinfo{person}{Caroline~M Johnston}, \bibinfo{person}{Simon
  Blessenohl}, {and} \bibinfo{person}{Phebe Vayanos}.}
  \bibinfo{year}{2020}\natexlab{}.
\newblock \showarticletitle{Preference Elicitation and Aggregation to Aid with
  Patient Triage during the COVID-19 Pandemic}.
\newblock \bibinfo{journal}{\emph{Workshop on Participatory Approaches to
  Machine Learning}} (\bibinfo{year}{2020}).
\newblock


\bibitem[\protect\citeauthoryear{Jones}{Jones}{2000}]%
        {casey2000neighborhood}
\bibfield{author}{\bibinfo{person}{Grant Jones}.}
  \bibinfo{year}{2000}\natexlab{}.
\newblock \bibinfo{title}{Developing a Neighborhood-Focused Agenda: Tools for
  Cities Getting Started}.
\newblock \bibinfo{howpublished}{The Annie E. Casey Foundation}.
\newblock


\bibitem[\protect\citeauthoryear{Jung, Kearns, Neel, Roth, Stapleton, and
  Wu}{Jung et~al\mbox{.}}{2021}]%
        {jung2021algorithmic}
\bibfield{author}{\bibinfo{person}{Christopher Jung}, \bibinfo{person}{Michael
  Kearns}, \bibinfo{person}{Seth Neel}, \bibinfo{person}{Aaron Roth},
  \bibinfo{person}{Logan Stapleton}, {and} \bibinfo{person}{Zhiwei~Steven Wu}.}
  \bibinfo{year}{2021}\natexlab{}.
\newblock \showarticletitle{An Algorithmic Framework for Fairness Elicitation}.
  In \bibinfo{booktitle}{\emph{Proceedings of the 2021 Symposium on the
  Foundations of Responsible Computing}}.
\newblock
\showeprint{https://arxiv.org/abs/1905.10660}


\bibitem[\protect\citeauthoryear{Kahng, Lee, Noothigattu, Procaccia, and
  Psomas}{Kahng et~al\mbox{.}}{2019}]%
        {kahng2019statistical}
\bibfield{author}{\bibinfo{person}{Anson Kahng}, \bibinfo{person}{Min~Kyung
  Lee}, \bibinfo{person}{Ritesh Noothigattu}, \bibinfo{person}{Ariel
  Procaccia}, {and} \bibinfo{person}{Christos-Alexandros Psomas}.}
  \bibinfo{year}{2019}\natexlab{}.
\newblock \showarticletitle{Statistical foundations of virtual democracy}. In
  \bibinfo{booktitle}{\emph{International Conference on Machine Learning}}.
  PMLR, \bibinfo{pages}{3173--3182}.
\newblock
\urldef\tempurl%
\url{https://proceedings.mlr.press/v97/kahng19a.html}
\showURL{%
\tempurl}


\bibitem[\protect\citeauthoryear{Kawakami, Sivaraman, Cheng, Stapleton, Cheng,
  Qing, Perer, Wu, Zhu, and Holstein}{Kawakami et~al\mbox{.}}{2022a}]%
        {kawakami2022partnerships}
\bibfield{author}{\bibinfo{person}{Anna Kawakami}, \bibinfo{person}{Venkatesh
  Sivaraman}, \bibinfo{person}{Hao-Fei Cheng}, \bibinfo{person}{Logan
  Stapleton}, \bibinfo{person}{Yanghuidi Cheng}, \bibinfo{person}{Diana Qing},
  \bibinfo{person}{Adam Perer}, \bibinfo{person}{Zhiwei~Steven Wu},
  \bibinfo{person}{Haiyi Zhu}, {and} \bibinfo{person}{Kenneth Holstein}.}
  \bibinfo{year}{2022}\natexlab{a}.
\newblock \showarticletitle{Improving Human-AI Partnerships in Child Welfare:
  Understanding Worker Practices, Challenges, and Desires for Algorithmic
  Decision Support}.
\newblock \bibinfo{journal}{\emph{Proceedings of the 2022 CHI Conference on
  Human Factors in Computing Systems}} (\bibinfo{year}{2022}).
\newblock
\urldef\tempurl%
\url{https://arxiv.org/abs/2204.02310}
\showURL{%
\tempurl}


\bibitem[\protect\citeauthoryear{Kawakami, Sivaraman, Stapleton, Cheng, Cheng,
  Qing, Perer, Wu, Zhu, and Holstein}{Kawakami et~al\mbox{.}}{2022b}]%
        {kawakami2022exploring}
\bibfield{author}{\bibinfo{person}{Anna Kawakami}, \bibinfo{person}{Venkatesh
  Sivaraman}, \bibinfo{person}{Logan Stapleton}, \bibinfo{person}{Hao-Fei
  Cheng}, \bibinfo{person}{Yanghuidi Cheng}, \bibinfo{person}{Diana Qing},
  \bibinfo{person}{Adam Perer}, \bibinfo{person}{Zhiwei~Steven Wu},
  \bibinfo{person}{Haiyi Zhu}, {and} \bibinfo{person}{Kenneth Holstein}.}
  \bibinfo{year}{2022}\natexlab{b}.
\newblock \showarticletitle{Exploring Interface Designs to Support AI-Assisted
  Decision Making in Child Welfare}.
\newblock \bibinfo{journal}{\emph{Proceedings of the 2022 ACM SIGCHI Conference
  on Designing Interactive Systems (DIS)}}.
\newblock


\bibitem[\protect\citeauthoryear{Kensing and Blomberg}{Kensing and
  Blomberg}{1998}]%
        {kensing1998participatory}
\bibfield{author}{\bibinfo{person}{F. Kensing} {and} \bibinfo{person}{J.
  Blomberg}.} \bibinfo{year}{1998}\natexlab{}.
\newblock \showarticletitle{Participatory Design: Issues and Concerns}.
\newblock \bibinfo{journal}{\emph{Computer Supported Cooperative Work (CSCW)}}
  \bibinfo{volume}{7} (\bibinfo{year}{1998}), \bibinfo{pages}{167---185}.
\newblock
\urldef\tempurl%
\url{https://doi.org/10.1023/A:1008689307411}
\showURL{%
\tempurl}


\bibitem[\protect\citeauthoryear{Kim, Wildeman, Jonson-Reid, and Drake}{Kim
  et~al\mbox{.}}{2017}]%
        {Kim2017lifetime}
\bibfield{author}{\bibinfo{person}{Hyunil Kim}, \bibinfo{person}{Christopher
  Wildeman}, \bibinfo{person}{Melissa Jonson-Reid}, {and}
  \bibinfo{person}{Brett Drake}.} \bibinfo{year}{2017}\natexlab{}.
\newblock \showarticletitle{Lifetime Prevalence of Investigating Child
  Maltreatment Among US Children}.
\newblock \bibinfo{journal}{\emph{American Journal of Public Health}}
  \bibinfo{volume}{107}, \bibinfo{number}{2} (\bibinfo{year}{2017}),
  \bibinfo{pages}{274--280}.
\newblock
\urldef\tempurl%
\url{https://doi.org/10.2105/AJPH.2016.303545}
\showDOI{\tempurl}


\bibitem[\protect\citeauthoryear{Kit}{Kit}{2021}]%
        {designkit2021}
\bibfield{author}{\bibinfo{person}{Google Design~Sprint Kit}.}
  \bibinfo{year}{2021}\natexlab{}.
\newblock \bibinfo{title}{Crazy 8’s Technique}.
\newblock
  \bibinfo{howpublished}{\url{https://designsprintkit.withgoogle.com/methodology/phase3-sketch/crazy-8s}}.
\newblock
\newblock
\shownote{Accessed: 2021-06.}


\bibitem[\protect\citeauthoryear{Knapp, Zeratsky, and Kowitz}{Knapp
  et~al\mbox{.}}{2016}]%
        {knapp2016sprint}
\bibfield{author}{\bibinfo{person}{Jake Knapp}, \bibinfo{person}{John
  Zeratsky}, {and} \bibinfo{person}{Braden Kowitz}.}
  \bibinfo{year}{2016}\natexlab{}.
\newblock \bibinfo{booktitle}{\emph{Sprint: How to solve big problems and test
  new ideas in just five days}}.
\newblock \bibinfo{publisher}{Simon and Schuster}.
\newblock


\bibitem[\protect\citeauthoryear{Knappenberger}{Knappenberger}{2020}]%
        {netflix2020gabrielfernandez}
\bibfield{author}{\bibinfo{person}{Brian Knappenberger}.}
  \bibinfo{year}{2020}\natexlab{}.
\newblock \bibinfo{title}{The Trials of Gabriel Fernandez: Improper Regard or
  Indifference}.
\newblock \bibinfo{howpublished}{Netflix}.
\newblock
\urldef\tempurl%
\url{https://www.netflix.com/watch/80234474}
\showURL{%
\tempurl}
\newblock
\shownote{Luminant Media and Common Sense Media (Producers).}


\bibitem[\protect\citeauthoryear{Kulynych, Madras, Milli, Raji, Zhou, and
  Zemel}{Kulynych et~al\mbox{.}}{2020}]%
        {paml}
\bibfield{author}{\bibinfo{person}{Bogdan Kulynych}, \bibinfo{person}{David
  Madras}, \bibinfo{person}{Smitha Milli}, \bibinfo{person}{Inioluwa~Deborah
  Raji}, \bibinfo{person}{Angela Zhou}, {and} \bibinfo{person}{Richard Zemel}.}
  \bibinfo{year}{2020}\natexlab{}.
\newblock \bibinfo{title}{Participatory Approaches to Machine Learning}.
\newblock \bibinfo{howpublished}{Internetional Conference on Machine Learning
  (ICML) Workshop}.
\newblock


\bibitem[\protect\citeauthoryear{Kyng and Mathiassen}{Kyng and
  Mathiassen}{1979}]%
        {kyng1979systems}
\bibfield{author}{\bibinfo{person}{Morten Kyng} {and} \bibinfo{person}{Lars
  Mathiassen}.} \bibinfo{year}{1979}\natexlab{}.
\newblock \bibinfo{booktitle}{\emph{Systems development and trade union
  activities}}.
\newblock \bibinfo{publisher}{Computer Science Department, Aarhus University
  Aarhus}.
\newblock


\bibitem[\protect\citeauthoryear{Larson, Angwin, Kirchner, and Mattu}{Larson
  et~al\mbox{.}}{2016}]%
        {propublica2016compas}
\bibfield{author}{\bibinfo{person}{Jeff Larson}, \bibinfo{person}{Julia
  Angwin}, \bibinfo{person}{Lauren Kirchner}, {and} \bibinfo{person}{Surya
  Mattu}.} \bibinfo{year}{2016}\natexlab{}.
\newblock \bibinfo{title}{How We Analyzed the COMPAS Recidivism Algorithm}.
\newblock
\newblock
\urldef\tempurl%
\url{https://www.propublica.org/article/how-we-analyzed-the-compas-recidivism-algorithm}
\showURL{%
\tempurl}


\bibitem[\protect\citeauthoryear{Lee, Kusbit, Kahng, Kim, Yuan, Chan, See,
  Noothigattu, Lee, Psomas, and Procaccia}{Lee et~al\mbox{.}}{2019}]%
        {lee2019webuildai}
\bibfield{author}{\bibinfo{person}{Min~Kyung Lee}, \bibinfo{person}{Daniel
  Kusbit}, \bibinfo{person}{Anson Kahng}, \bibinfo{person}{Ji~Tae Kim},
  \bibinfo{person}{Xinran Yuan}, \bibinfo{person}{Allissa Chan},
  \bibinfo{person}{Daniel See}, \bibinfo{person}{Ritesh Noothigattu},
  \bibinfo{person}{Siheon Lee}, \bibinfo{person}{Alexandros Psomas}, {and}
  \bibinfo{person}{Ariel~D. Procaccia}.} \bibinfo{year}{2019}\natexlab{}.
\newblock \showarticletitle{WeBuildAI: Participatory Framework for Algorithmic
  Governance}.
\newblock \bibinfo{journal}{\emph{Proceedings of the ACM on Human Computer
  Interaction}} \bibinfo{volume}{3}, \bibinfo{number}{CSCW}
  (\bibinfo{date}{nov} \bibinfo{year}{2019}).
\newblock
\urldef\tempurl%
\url{https://doi.org/10.1145/3359283}
\showURL{%
\tempurl}


\bibitem[\protect\citeauthoryear{Loi, Lodato, Wolf, Arar, and Blomberg}{Loi
  et~al\mbox{.}}{2018}]%
        {loi2018pd}
\bibfield{author}{\bibinfo{person}{Daria Loi}, \bibinfo{person}{Thomas Lodato},
  \bibinfo{person}{Christine~T. Wolf}, \bibinfo{person}{Raphael Arar}, {and}
  \bibinfo{person}{Jeanette Blomberg}.} \bibinfo{year}{2018}\natexlab{}.
\newblock \showarticletitle{PD Manifesto for AI Futures}. In
  \bibinfo{booktitle}{\emph{Proceedings of the 15th Participatory Design
  Conference: Short Papers, Situated Actions, Workshops and Tutorial - Volume
  2}} (Hasselt and Genk, Belgium) \emph{(\bibinfo{series}{PDC '18})}.
  \bibinfo{publisher}{Association for Computing Machinery},
  \bibinfo{address}{New York, NY, USA}, Article \bibinfo{articleno}{48},
  \bibinfo{numpages}{4}~pages.
\newblock
\showISBNx{9781450355742}
\urldef\tempurl%
\url{https://doi.org/10.1145/3210604.3210614}
\showDOI{\tempurl}


\bibitem[\protect\citeauthoryear{Mickelson, LaLiberte, and Piescher}{Mickelson
  et~al\mbox{.}}{2017}]%
        {mickelson2017assessing}
\bibfield{author}{\bibinfo{person}{Nicole Mickelson}, \bibinfo{person}{Traci
  LaLiberte}, {and} \bibinfo{person}{Kristine Piescher}.}
  \bibinfo{year}{2017}\natexlab{}.
\newblock \showarticletitle{Assessing risk: A comparison of tools for child
  welfare practice with indigenous families}.
\newblock \bibinfo{journal}{\emph{Center for Advanced Studies in Child Welfare,
  University of Minnesota, St. Paul, MN}} (\bibinfo{year}{2017}).
\newblock
\urldef\tempurl%
\url{https://cascw.umn.edu/wp-content/uploads/2018/01/Risk-Assessment_FinalReport.pdf}
\showURL{%
\tempurl}


\bibitem[\protect\citeauthoryear{Muller and Kuhn}{Muller and Kuhn}{1993}]%
        {muller1993participatory}
\bibfield{author}{\bibinfo{person}{Michael~J Muller} {and}
  \bibinfo{person}{Sarah Kuhn}.} \bibinfo{year}{1993}\natexlab{}.
\newblock \showarticletitle{Participatory design}.
\newblock \bibinfo{journal}{\emph{Commun. ACM}} \bibinfo{volume}{36},
  \bibinfo{number}{6} (\bibinfo{year}{1993}), \bibinfo{pages}{24--28}.
\newblock


\bibitem[\protect\citeauthoryear{Nash}{Nash}{2017}]%
        {nash2017losangeles}
\bibfield{author}{\bibinfo{person}{Judge~Michael Nash}.}
  \bibinfo{year}{2017}\natexlab{}.
\newblock \bibinfo{title}{EXAMINATION OF USING STRUCTURED DECISION MAKING AND
  PREDICTIVE ANALYTICS IN ASSESSING SAFETY AND RISK IN CHILD WELFARE (ITEM NO.
  49-A, AGENDA OF SEPTEMBER 20, 2016)}.
\newblock
\newblock
\urldef\tempurl%
\url{http://ocp.lacounty.gov/Portals/OCP/PDF/Reports\%20and\%20Communication/Safety\%20and\%20Risk\%20Assessment/SDM\%20and\%20Predictive\%20Analytics\%20Report\%20(Risk\%20Assessment\%20Tools)\%20(May\%202017).pdf?ver=2018-10-24-083910-100}
\showURL{%
\tempurl}
\newblock
\shownote{Online; accessed 6-January-2022.}


\bibitem[\protect\citeauthoryear{National Coalition~for Child
  Protection~Reform}{National Coalition~for Child Protection~Reform}{2017}]%
        {nccpr2017losangeles}
\bibfield{author}{\bibinfo{person}{NCCPR: National Coalition~for Child
  Protection~Reform}.} \bibinfo{year}{2017}\natexlab{}.
\newblock \bibinfo{title}{Los Angeles County quietly drops its first child
  welfare predictive analytics experiment}.
\newblock
\newblock
\urldef\tempurl%
\url{https://www.nccprblog.org/2017/05/los-angeles-county-quietly-drops-its.html}
\showURL{%
\tempurl}
\newblock
\shownote{Online; accessed 6-January-2022.}


\bibitem[\protect\citeauthoryear{National Coalition~for Child
  Protection~Reform}{National Coalition~for Child Protection~Reform}{2018}]%
        {nccpr2018predictive}
\bibfield{author}{\bibinfo{person}{NCCPR: National Coalition~for Child
  Protection~Reform}.} \bibinfo{year}{2018}\natexlab{}.
\newblock \bibinfo{title}{Predictive analytics in Pittsburgh child welfare: Was
  the ``ethics review'' of Allegheny County's ``scarlet number'' algorithm
  ethical?}
\newblock
\newblock
\urldef\tempurl%
\url{https://www.nccprblog.org/2018/03/predictive-analytics-in-pittsburgh.html}
\showURL{%
\tempurl}
\newblock
\shownote{Online; accessed 8-September-2021.}


\bibitem[\protect\citeauthoryear{National Coalition~for Child
  Protection~Reform}{National Coalition~for Child Protection~Reform}{2019}]%
        {nccpr2019racial}
\bibfield{author}{\bibinfo{person}{NCCPR: National Coalition~for Child
  Protection~Reform}.} \bibinfo{year}{2019}\natexlab{}.
\newblock \bibinfo{title}{No, you can’t use predictive analytics to reduce
  racial bias in child welfare}.
\newblock
\newblock
\urldef\tempurl%
\url{https://www.nccprblog.org/2019/06/no-you-cant-use-predictive-analytics-to.html}
\showURL{%
\tempurl}
\newblock
\shownote{Online; accessed 2-December-2021.}


\bibitem[\protect\citeauthoryear{Noothigattu, Gaikwad, Awad, Dsouza, Rahwan,
  Ravikumar, and Procaccia}{Noothigattu et~al\mbox{.}}{2018}]%
        {noothigattu2018voting}
\bibfield{author}{\bibinfo{person}{Ritesh Noothigattu},
  \bibinfo{person}{Snehalkumar Gaikwad}, \bibinfo{person}{Edmond Awad},
  \bibinfo{person}{Sohan Dsouza}, \bibinfo{person}{Iyad Rahwan},
  \bibinfo{person}{Pradeep Ravikumar}, {and} \bibinfo{person}{Ariel
  Procaccia}.} \bibinfo{year}{2018}\natexlab{}.
\newblock \showarticletitle{A voting-based system for ethical decision making}.
  In \bibinfo{booktitle}{\emph{Proceedings of the AAAI Conference on Artificial
  Intelligence}}, Vol.~\bibinfo{volume}{32}.
\newblock
\urldef\tempurl%
\url{https://arxiv.org/abs/1709.06692}
\showURL{%
\tempurl}


\bibitem[\protect\citeauthoryear{of~Human~Services}{of~Human~Services}{[n.d.]a}]%
        {afstfaq}
\bibfield{author}{\bibinfo{person}{Allegheny County~Department of
  Human~Services}.} \bibinfo{year}{[n.d.]}\natexlab{a}.
\newblock \bibinfo{title}{Allegheny Family Screening Tool, Frequently-Asked
  Questions | Updated August 2018}.
\newblock
  \bibinfo{howpublished}{\url{https://www.alleghenycountyanalytics.us/wp-content/uploads/2018/10/17-ACDHS-11_AFST_102518.pdf}}.
\newblock
\newblock
\shownote{Online; accessed 8-September-2021.}


\bibitem[\protect\citeauthoryear{of~Human~Services}{of~Human~Services}{[n.d.]b}]%
        {hellobabyfaq}
\bibfield{author}{\bibinfo{person}{Allegheny County~Department of
  Human~Services}.} \bibinfo{year}{[n.d.]}\natexlab{b}.
\newblock \bibinfo{title}{Hello Baby Frequently Asked Questions}.
\newblock
  \bibinfo{howpublished}{\url{https://www.alleghenycountyanalytics.us/wp-content/uploads/2020/11/HB_FAQ-updated-11-5-2020.pdf}}.
\newblock
\newblock
\shownote{Online; accessed 20-January-2022.}


\bibitem[\protect\citeauthoryear{of~Human~Services}{of~Human~Services}{2019}]%
        {dhs2019impactsummary}
\bibfield{author}{\bibinfo{person}{Allegheny County~Department of
  Human~Services}.} \bibinfo{year}{2019}\natexlab{}.
\newblock \showarticletitle{Impact Evaluation Summary of the Allegheny Family
  Screening Tool}.
\newblock \bibinfo{journal}{\emph{Pittsburgh: Allegheny County}}
  (\bibinfo{year}{2019}).
\newblock
\urldef\tempurl%
\url{https://www.alleghenycountyanalytics.us/wp-content/uploads/2019/05/Impact-Evaluation-Summary-from-16-ACDHS-26_PredictiveRisk_Package_050119_FINAL-5.pdf}
\showURL{%
\tempurl}


\bibitem[\protect\citeauthoryear{Ogbonnaya-Ogburu, Smith, To, and
  Toyama}{Ogbonnaya-Ogburu et~al\mbox{.}}{2020}]%
        {ogbonnaya-ogburu2020critical}
\bibfield{author}{\bibinfo{person}{Ihudiya~Finda Ogbonnaya-Ogburu},
  \bibinfo{person}{Angela~D.R. Smith}, \bibinfo{person}{Alexandra To}, {and}
  \bibinfo{person}{Kentaro Toyama}.} \bibinfo{year}{2020}\natexlab{}.
\newblock \showarticletitle{Critical Race Theory for HCI}. In
  \bibinfo{booktitle}{\emph{Proceedings of the 2020 CHI Conference on Human
  Factors in Computing Systems (CHI'20)}}. \bibinfo{pages}{1---16}.
\newblock
\urldef\tempurl%
\url{https://doi.org/10.1145/3313831.3376392}
\showURL{%
\tempurl}


\bibitem[\protect\citeauthoryear{on~Crime and Center}{on~Crime and
  Center}{[n.d.]}]%
        {sdm}
\bibfield{author}{\bibinfo{person}{National~Council on Crime} {and}
  \bibinfo{person}{Delinquency Children’s~Research Center}.}
  \bibinfo{year}{[n.d.]}\natexlab{}.
\newblock \bibinfo{title}{The Structured Decision Making System for Child
  Protective Services Policy and Procedures Manual}.
\newblock
  \bibinfo{howpublished}{\url{https://www.cdss.ca.gov/Portals/9/SDM\%20Policy\%20and\%20Procedure\%20Manual.pdf}}.
\newblock
\newblock
\shownote{Online; accessed 8-September-2021.}


\bibitem[\protect\citeauthoryear{(PAIR)}{(PAIR)}{2020}]%
        {pair2020boundary}
\bibfield{author}{\bibinfo{person}{Google People + AI~Research (PAIR)}.}
  \bibinfo{year}{2020}\natexlab{}.
\newblock \bibinfo{title}{Boundary Objects for Participatory Machine Learning}.
\newblock \bibinfo{howpublished}{\url{https://youtu.be/J2smNXwflYs}}.
\newblock
\newblock
\shownote{Last Accessed: May 10, 2022.}


\bibitem[\protect\citeauthoryear{Palacin, Nelimarkka, Reynolds-Cu\'{e}llar, and
  Becker}{Palacin et~al\mbox{.}}{2020}]%
        {palacin2020pseudoparticipation}
\bibfield{author}{\bibinfo{person}{Victoria Palacin}, \bibinfo{person}{Matti
  Nelimarkka}, \bibinfo{person}{Pedro Reynolds-Cu\'{e}llar}, {and}
  \bibinfo{person}{Christoph Becker}.} \bibinfo{year}{2020}\natexlab{}.
\newblock \showarticletitle{The Design of Pseudo-Participation}. In
  \bibinfo{booktitle}{\emph{Proceedings of the 16th Participatory Design
  Conference 2020 - Participation(s) Otherwise - Volume 2}}
  \emph{(\bibinfo{series}{PDC '20})}. \bibinfo{publisher}{Association for
  Computing Machinery}, \bibinfo{address}{New York, NY, USA},
  \bibinfo{pages}{40–44}.
\newblock
\showISBNx{9781450376068}
\urldef\tempurl%
\url{https://doi.org/10.1145/3384772.3385141}
\showDOI{\tempurl}


\bibitem[\protect\citeauthoryear{{Panoptykon Foundation}}{{Panoptykon
  Foundation}}{2015}]%
        {panoptykon2015unemployed}
\bibfield{author}{\bibinfo{person}{{Panoptykon Foundation}}.}
  \bibinfo{year}{2015}\natexlab{}.
\newblock \bibinfo{title}{Profiling the unemployed in Poland: Social and
  political implications of algorithmic decision-making}.
\newblock
\newblock


\bibitem[\protect\citeauthoryear{Pelton}{Pelton}{1994}]%
        {pelton1994}
\bibfield{author}{\bibinfo{person}{Leroy Pelton}.}
  \bibinfo{year}{1994}\natexlab{}.
\newblock \bibinfo{booktitle}{\emph{Has Permanency Planning Been Successful?
  No}}.
\newblock


\bibitem[\protect\citeauthoryear{Pierre, Crooks, Currie, Paris, and
  Pasquetto}{Pierre et~al\mbox{.}}{2021}]%
        {pierre2021getting}
\bibfield{author}{\bibinfo{person}{Jennifer Pierre}, \bibinfo{person}{Roderic
  Crooks}, \bibinfo{person}{Morgan Currie}, \bibinfo{person}{Britt Paris},
  {and} \bibinfo{person}{Irene Pasquetto}.} \bibinfo{year}{2021}\natexlab{}.
\newblock \showarticletitle{Getting Ourselves Together: Data-centered
  participatory design research \& epistemic burden}. In
  \bibinfo{booktitle}{\emph{Proceedings of the 2021 CHI Conference on Human
  Factors in Computing Systems}}. \bibinfo{pages}{1--11}.
\newblock


\bibitem[\protect\citeauthoryear{Pomeroy}{Pomeroy}{2019}]%
        {pomeroy2019community}
\bibfield{author}{\bibinfo{person}{Carrie Pomeroy}.}
  \bibinfo{year}{2019}\natexlab{}.
\newblock \bibinfo{title}{How community members in Ramsey County stopped a
  big-data plan from flagging students as at-risk}.
\newblock
\newblock
\urldef\tempurl%
\url{https://www.tcdailyplanet.net/how-community-members-in-ramsey-county-stopped-a-big-data-plan-from-flagging-students-as-at-risk/}
\showURL{%
\tempurl}


\bibitem[\protect\citeauthoryear{Posner, Brent, Lucas, Gould, Stanley, Brown,
  Fisher, Zelazny, Burke, Oquendo, and Mann}{Posner et~al\mbox{.}}{2008}]%
        {posner2008columbiasuicide}
\bibfield{author}{\bibinfo{person}{K. Posner}, \bibinfo{person}{D. Brent},
  \bibinfo{person}{C. Lucas}, \bibinfo{person}{M. Gould}, \bibinfo{person}{B.
  Stanley}, \bibinfo{person}{G. Brown}, \bibinfo{person}{P. Fisher},
  \bibinfo{person}{J. Zelazny}, \bibinfo{person}{A. Burke}, \bibinfo{person}{M.
  Oquendo}, {and} \bibinfo{person}{J. Mann}.} \bibinfo{year}{2008}\natexlab{}.
\newblock \showarticletitle{Columbia-suicide severity rating scale (C-SSRS)}.
\newblock \bibinfo{journal}{\emph{Columbia University Medical Center}}
  (\bibinfo{year}{2008}).
\newblock
\showeprint{https://vtspc.org/wp-content/uploads/2016/12/C-SSRS-LifetimeRecent-Clinical.pdf}


\bibitem[\protect\citeauthoryear{Posner, Brown, Stanley, Brent, Yershova,
  Oquendo, Currier, Melvin, Greenhill, Shen, and Mann}{Posner
  et~al\mbox{.}}{2011}]%
        {posner2011columbiasuicide}
\bibfield{author}{\bibinfo{person}{Kelly Posner}, \bibinfo{person}{Gregory~K.
  Brown}, \bibinfo{person}{Barbara Stanley}, \bibinfo{person}{David~A. Brent},
  \bibinfo{person}{Kseniya~V. Yershova}, \bibinfo{person}{Maria~A. Oquendo},
  \bibinfo{person}{Glenn~W. Currier}, \bibinfo{person}{Glenn~A. Melvin},
  \bibinfo{person}{Laurence Greenhill}, \bibinfo{person}{Sa Shen}, {and}
  \bibinfo{person}{J.~John Mann}.} \bibinfo{year}{2011}\natexlab{}.
\newblock \showarticletitle{The Columbia–Suicide Severity Rating Scale:
  Initial Validity and Internal Consistency Findings From Three Multisite
  Studies With Adolescents and Adults}.
\newblock \bibinfo{journal}{\emph{American Journal of Psychiatry}}
  \bibinfo{volume}{168}, \bibinfo{number}{12} (\bibinfo{year}{2011}),
  \bibinfo{pages}{1266--1277}.
\newblock
\urldef\tempurl%
\url{https://doi.org/10.1176/appi.ajp.2011.10111704}
\showURL{%
\tempurl}


\bibitem[\protect\citeauthoryear{Power}{Power}{[n.d.]}]%
        {familypower}
\bibfield{author}{\bibinfo{person}{Movement For~Family Power}.}
  \bibinfo{year}{[n.d.]}\natexlab{}.
\newblock \bibinfo{title}{Movement For Family Power}.
\newblock
  \bibinfo{howpublished}{\url{https://www.movementforfamilypower.org/}}.
\newblock
\newblock
\shownote{Last Accessed: May 10, 2022.}


\bibitem[\protect\citeauthoryear{Programs}{Programs}{2021}]%
        {casey2021prevention}
\bibfield{author}{\bibinfo{person}{Casey~Family Programs}.}
  \bibinfo{year}{2021}\natexlab{}.
\newblock \bibinfo{title}{Strategy Brief: How do parent partner programs
  instill hope and support prevention and reunification?}
\newblock
  \bibinfo{howpublished}{\url{https://caseyfamilypro-wpengine.netdna-ssl.com/media/HO_Parent-Partner-Program_01-21.pdf}}.
\newblock
\newblock
\shownote{Accessed: 01-20-2022.}


\bibitem[\protect\citeauthoryear{Riley}{Riley}{2018}]%
        {riley2018can}
\bibfield{author}{\bibinfo{person}{Naomi~Schaefer Riley}.}
  \bibinfo{year}{2018}\natexlab{}.
\newblock \bibinfo{title}{Can Big Data Help Save Abused Kids?}
\newblock
\newblock
\urldef\tempurl%
\url{https://reason.com/2018/01/22/can-big-data-help-save-abused/}
\showURL{%
\tempurl}
\newblock
\shownote{Online; accessed 8-September-2021.}


\bibitem[\protect\citeauthoryear{Rise}{Rise}{[n.d.]}]%
        {rise}
\bibfield{author}{\bibinfo{person}{Rise}.} \bibinfo{year}{[n.d.]}\natexlab{}.
\newblock \bibinfo{title}{Rise}.
\newblock \bibinfo{howpublished}{\url{https://www.risemagazine.org}}.
\newblock
\newblock
\shownote{Last Accessed: May 10, 2022.}


\bibitem[\protect\citeauthoryear{Roberts}{Roberts}{2002}]%
        {roberts2002shattered}
\bibfield{author}{\bibinfo{person}{Dorothy~E. Roberts}.}
  \bibinfo{year}{2002}\natexlab{}.
\newblock \bibinfo{booktitle}{\emph{Shattered bonds: The color of child
  welfare}}.
\newblock \bibinfo{publisher}{Basic Books}, \bibinfo{address}{New York}.
\newblock


\bibitem[\protect\citeauthoryear{Roberts}{Roberts}{2005}]%
        {roberts2005community}
\bibfield{author}{\bibinfo{person}{Dorothy~E Roberts}.}
  \bibinfo{year}{2005}\natexlab{}.
\newblock \showarticletitle{The Community Dimension of State Child Protection}.
\newblock \bibinfo{journal}{\emph{Hofstra L. Rev.}}  \bibinfo{volume}{34}
  (\bibinfo{year}{2005}), \bibinfo{pages}{23}.
\newblock


\bibitem[\protect\citeauthoryear{Roberts}{Roberts}{2007}]%
        {roberts2007paradox}
\bibfield{author}{\bibinfo{person}{Dorothy~E. Roberts}.}
  \bibinfo{year}{2007}\natexlab{}.
\newblock \showarticletitle{Child Welfare's Paradox}.
\newblock \bibinfo{journal}{\emph{William \& Mary Law Review}}
  (\bibinfo{year}{2007}).
\newblock
\urldef\tempurl%
\url{https://scholarship.law.upenn.edu/faculty_scholarship/578}
\showURL{%
\tempurl}


\bibitem[\protect\citeauthoryear{Roberts}{Roberts}{2008}]%
        {roberts2008paradox}
\bibfield{author}{\bibinfo{person}{Dorothy~E. Roberts}.}
  \bibinfo{year}{2008}\natexlab{}.
\newblock \showarticletitle{The racial geography of child welfare: toward a new
  research paradigm}.
\newblock \bibinfo{journal}{\emph{Child welfare}}  \bibinfo{volume}{87}
  (\bibinfo{year}{2008}).
\newblock


\bibitem[\protect\citeauthoryear{Roberts}{Roberts}{2019}]%
        {roberts2019digitizing}
\bibfield{author}{\bibinfo{person}{Dorothy~E. Roberts}.}
  \bibinfo{year}{2019}\natexlab{}.
\newblock \showarticletitle{Digitizing the Carceral State (Review of
  \cite{eubanks2018automating})}.
\newblock \bibinfo{journal}{\emph{Harvard Law Review}}  \bibinfo{volume}{132}
  (\bibinfo{year}{2019}), \bibinfo{pages}{1695---1728}.
\newblock


\bibitem[\protect\citeauthoryear{Roberts}{Roberts}{2022}]%
        {roberts2022torn}
\bibfield{author}{\bibinfo{person}{Dorothy~E. Roberts}.}
  \bibinfo{year}{2022}\natexlab{}.
\newblock \bibinfo{booktitle}{\emph{Torn Apart: How the Child Welfare System
  Destroys Black Families--and How Abolition Can Build a Safer World}}.
\newblock \bibinfo{publisher}{Basic Books}.
\newblock
\showISBNx{9781541675452}


\bibitem[\protect\citeauthoryear{Robertson and Salehi}{Robertson and
  Salehi}{2020}]%
        {robertson2020if}
\bibfield{author}{\bibinfo{person}{Samantha Robertson} {and}
  \bibinfo{person}{Niloufar Salehi}.} \bibinfo{year}{2020}\natexlab{}.
\newblock \showarticletitle{What If I Don't Like Any Of The Choices? The Limits
  of Preference Elicitation for Participatory Algorithm Design}.
\newblock \bibinfo{journal}{\emph{arXiv preprint arXiv:2007.06718}}
  (\bibinfo{year}{2020}).
\newblock


\bibitem[\protect\citeauthoryear{Samant, Horowitz, Xu, and Beiers}{Samant
  et~al\mbox{.}}{2021}]%
        {aclu2021family}
\bibfield{author}{\bibinfo{person}{Anjana Samant}, \bibinfo{person}{Aaron
  Horowitz}, \bibinfo{person}{Kath Xu}, {and} \bibinfo{person}{Sophie Beiers}.}
  \bibinfo{year}{2021}\natexlab{}.
\newblock \showarticletitle{Family Surveillance by Algorithm: The Rapidly
  Spreading Tools Few Have Heard Of}.
\newblock \bibinfo{journal}{\emph{American Civil Liberties Union (ACLU)}}
  (\bibinfo{year}{2021}).
\newblock
\urldef\tempurl%
\url{https://www.aclu.org/sites/default/files/field_document/2021.09.28a_family_surveillance_by_algorithm.pdf}
\showURL{%
\tempurl}


\bibitem[\protect\citeauthoryear{Sandberg}{Sandberg}{1979}]%
        {sandberg1979computers}
\bibfield{author}{\bibinfo{person}{{\AA}ke Sandberg}.}
  \bibinfo{year}{1979}\natexlab{}.
\newblock \bibinfo{booktitle}{\emph{Computers dividing man and work: Recent
  Scandinavian research on planning and computers from a trade union
  perspective}}.
\newblock \bibinfo{publisher}{Arbetslivcentrum}.
\newblock


\bibitem[\protect\citeauthoryear{Sapien}{Sapien}{2016}]%
        {sapien2016foil}
\bibfield{author}{\bibinfo{person}{Joaquin Sapien}.}
  \bibinfo{year}{2016}\natexlab{}.
\newblock \showarticletitle{Foiled by FOIL: How One City Agency Has Dragged Out
  a Request for Public Records for Nearly a Year}.
\newblock \bibinfo{journal}{\emph{ProPublica}} (\bibinfo{date}{April}
  \bibinfo{year}{2016}).
\newblock
\urldef\tempurl%
\url{https://www.propublica.org/article/how-city-agency-dragged-out-request-for-public-records-for-nearly-a-year}
\showURL{%
\tempurl}


\bibitem[\protect\citeauthoryear{Saxena, Badillo-Urquiola, Wisniewski, and
  Guha}{Saxena et~al\mbox{.}}{2020}]%
        {saxena2020human}
\bibfield{author}{\bibinfo{person}{Devansh Saxena}, \bibinfo{person}{Karla
  Badillo-Urquiola}, \bibinfo{person}{Pamela~J Wisniewski}, {and}
  \bibinfo{person}{Shion Guha}.} \bibinfo{year}{2020}\natexlab{}.
\newblock \showarticletitle{A Human-Centered Review of Algorithms used within
  the US Child Welfare System}. In \bibinfo{booktitle}{\emph{Proceedings of the
  2020 CHI Conference on Human Factors in Computing Systems}}.
  \bibinfo{pages}{1--15}.
\newblock


\bibitem[\protect\citeauthoryear{Saxena, Badillo-Urquiola, Wisniewski, and
  Guha}{Saxena et~al\mbox{.}}{2021}]%
        {saxena2021framework}
\bibfield{author}{\bibinfo{person}{Devansh Saxena}, \bibinfo{person}{Karla
  Badillo-Urquiola}, \bibinfo{person}{Pamela~J. Wisniewski}, {and}
  \bibinfo{person}{Shion Guha}.} \bibinfo{year}{2021}\natexlab{}.
\newblock \showarticletitle{A Framework of High-Stakes Algorithmic
  Decision-Making for the Public Sector Developed through a Case Study of
  Child-Welfare}.
\newblock  \bibinfo{volume}{5}, \bibinfo{number}{CSCW2} (\bibinfo{year}{2021}),
  \bibinfo{numpages}{41}~pages.
\newblock
\urldef\tempurl%
\url{https://doi.org/10.1145/3476089}
\showDOI{\tempurl}


\bibitem[\protect\citeauthoryear{Saxena and Guha}{Saxena and Guha}{2020}]%
        {saxena2020participatory}
\bibfield{author}{\bibinfo{person}{Devansh Saxena} {and} \bibinfo{person}{Shion
  Guha}.} \bibinfo{year}{2020}\natexlab{}.
\newblock \showarticletitle{Conducting Participatory Design to Improve
  Algorithms in Public Services: Lessons and Challenges}. In
  \bibinfo{booktitle}{\emph{Conference Companion Publication of the 2020 on
  Computer Supported Cooperative Work and Social Computing}}.
  \bibinfo{pages}{383–--388}.
\newblock


\bibitem[\protect\citeauthoryear{Saxena, Repaci, Sage, and Guha}{Saxena
  et~al\mbox{.}}{2022}]%
        {saxena2022how}
\bibfield{author}{\bibinfo{person}{Devansh Saxena}, \bibinfo{person}{Charles
  Repaci}, \bibinfo{person}{Melanie~D Sage}, {and} \bibinfo{person}{Shion
  Guha}.} \bibinfo{year}{2022}\natexlab{}.
\newblock \showarticletitle{How to Train a (Bad) Algorithmic Caseworker: A
  Quantitative Deconstruction of Risk Assessments in Child Welfare}. In
  \bibinfo{booktitle}{\emph{CHI Conference on Human Factors in Computing
  Systems Extended Abstracts}} (New Orleans, LA, USA)
  \emph{(\bibinfo{series}{CHI EA '22})}. \bibinfo{publisher}{Association for
  Computing Machinery}, \bibinfo{address}{New York, NY, USA}, Article
  \bibinfo{articleno}{407}, \bibinfo{numpages}{7}~pages.
\newblock
\showISBNx{9781450391566}
\urldef\tempurl%
\url{https://doi.org/10.1145/3491101.3519771}
\showDOI{\tempurl}


\bibitem[\protect\citeauthoryear{Scott, Wang, Miceli, Delobelle,
  Sztandar-Sztanderska, and Berendt}{Scott et~al\mbox{.}}{2022}]%
        {scott2022algorithmic}
\bibfield{author}{\bibinfo{person}{Kristen Scott}, \bibinfo{person}{Sonja~Mei
  Wang}, \bibinfo{person}{Milagros Miceli}, \bibinfo{person}{Pieter Delobelle},
  \bibinfo{person}{Karolina Sztandar-Sztanderska}, {and}
  \bibinfo{person}{Bettina Berendt}.} \bibinfo{year}{2022}\natexlab{}.
\newblock \showarticletitle{Algorithmic Tools in Public Employment Services:
  Towards a Jobseeker-Centric Perspective}. In
  \bibinfo{booktitle}{\emph{Proceedings of 2022 Conference on Fairness,
  Accountability, and Transparency (FAccT) (forthcoming)}}.
  \bibinfo{publisher}{ACM}.
\newblock


\bibitem[\protect\citeauthoryear{Sloane, Moss, Awomolo, and Forlano}{Sloane
  et~al\mbox{.}}{2020}]%
        {sloane2020participation}
\bibfield{author}{\bibinfo{person}{Mona Sloane}, \bibinfo{person}{Emanuel
  Moss}, \bibinfo{person}{Olaitan Awomolo}, {and} \bibinfo{person}{Laura
  Forlano}.} \bibinfo{year}{2020}\natexlab{}.
\newblock \showarticletitle{Participation is not a Design Fix for Machine
  Learning}.
\newblock \bibinfo{journal}{\emph{arXiv preprint}}
  \bibinfo{volume}{arXiv:2007.02423} (\bibinfo{year}{2020}).
\newblock


\bibitem[\protect\citeauthoryear{Smith, Yu, Srivastava, Halfaker, Terveen, and
  Zhu}{Smith et~al\mbox{.}}{2020}]%
        {smith2020community}
\bibfield{author}{\bibinfo{person}{C.~Estelle Smith}, \bibinfo{person}{Bowen
  Yu}, \bibinfo{person}{Anjali Srivastava}, \bibinfo{person}{Aaron Halfaker},
  \bibinfo{person}{Loren Terveen}, {and} \bibinfo{person}{Haiyi Zhu}.}
  \bibinfo{year}{2020}\natexlab{}.
\newblock \bibinfo{booktitle}{\emph{Keeping Community in the Loop:
  Understanding Wikipedia Stakeholder Values for Machine Learning-Based
  Systems}}.
\newblock \bibinfo{pages}{1–14}.
\newblock


\bibitem[\protect\citeauthoryear{Spinuzzi}{Spinuzzi}{2002}]%
        {spinuzzi2002scandinavian}
\bibfield{author}{\bibinfo{person}{Clay Spinuzzi}.}
  \bibinfo{year}{2002}\natexlab{}.
\newblock \showarticletitle{A Scandinavian Challenge, a US Response:
  Methodological Assumptions in Scandinavian and US Prototyping Approaches}. In
  \bibinfo{booktitle}{\emph{Proceedings of the 20th Annual International
  Conference on Computer Documentation}} \emph{(\bibinfo{series}{SIGDOC '02})}.
  \bibinfo{publisher}{Association for Computing Machinery},
  \bibinfo{address}{New York, NY, USA}, \bibinfo{pages}{208–215}.
\newblock
\showISBNx{1581135432}
\urldef\tempurl%
\url{https://doi.org/10.1145/584955.584986}
\showDOI{\tempurl}


\bibitem[\protect\citeauthoryear{Stack}{Stack}{2018}]%
        {stack2018cyf}
\bibfield{author}{\bibinfo{person}{Barbara~White Stack}.}
  \bibinfo{year}{2018}\natexlab{}.
\newblock \showarticletitle{CYF: An agency that works, helping kids and their
  families}.
\newblock \bibinfo{journal}{\emph{Pittsburgh Post-Gazette}}
  (\bibinfo{year}{2018}).
\newblock
\urldef\tempurl%
\url{https://www.post-gazette.com/opinion/Op-Ed/2018/02/11/CYF-An-agency-that-works-helping-kids-and-their-families/stories/201802040037}
\showURL{%
\tempurl}
\newblock
\shownote{Online; accessed 8-September-2021.}


\bibitem[\protect\citeauthoryear{Stevenson}{Stevenson}{2018}]%
        {stevenson2018assessing}
\bibfield{author}{\bibinfo{person}{Megan Stevenson}.}
  \bibinfo{year}{2018}\natexlab{}.
\newblock \showarticletitle{Assessing Risk Assessment in Action}.
\newblock \bibinfo{journal}{\emph{Law, Economics, and Business Fellows’
  Discussion Paper Series 85}}  \bibinfo{volume}{103} (\bibinfo{year}{2018}),
  \bibinfo{pages}{303--384}.
\newblock
\urldef\tempurl%
\url{https://scholarship.law.umn.edu/mlr/58/}
\showURL{%
\tempurl}


\bibitem[\protect\citeauthoryear{Stevenson and Doleac}{Stevenson and
  Doleac}{2021}]%
        {stevenson2021algorithmic}
\bibfield{author}{\bibinfo{person}{Megan Stevenson} {and}
  \bibinfo{person}{Jennifer Doleac}.} \bibinfo{year}{2021}\natexlab{}.
\newblock \showarticletitle{Algorithmic Risk Assessment in the Hands of
  Humans}.
\newblock  (\bibinfo{year}{2021}).
\newblock
\urldef\tempurl%
\url{https://papers.ssrn.com/sol3/papers.cfm?abstract_id=3489440}
\showURL{%
\tempurl}


\bibitem[\protect\citeauthoryear{Technology}{Technology}{[n.d.]}]%
        {mindshare}
\bibfield{author}{\bibinfo{person}{MindShare Technology}.}
  \bibinfo{year}{[n.d.]}\natexlab{}.
\newblock \bibinfo{title}{Improving Outcomes using data you already have}.
\newblock
  \bibinfo{howpublished}{\url{https://mindshare-technology.com/analytics/}}.
\newblock
\newblock
\shownote{Online; accessed 8-September-2021.}


\bibitem[\protect\citeauthoryear{Testa and Kelly}{Testa and Kelly}{2020}]%
        {testa2020evolution}
\bibfield{author}{\bibinfo{person}{Mark~F Testa} {and} \bibinfo{person}{David
  Kelly}.} \bibinfo{year}{2020}\natexlab{}.
\newblock \showarticletitle{The evolution of federal child welfare policy
  through the Family First Prevention Services Act of 2018: Opportunities,
  barriers, and unintended consequences}.
\newblock \bibinfo{journal}{\emph{The ANNALS of the American Academy of
  Political and Social Science}} \bibinfo{volume}{692}, \bibinfo{number}{1}
  (\bibinfo{year}{2020}), \bibinfo{pages}{68--96}.
\newblock


\bibitem[\protect\citeauthoryear{Toros and Flaming}{Toros and Flaming}{2018}]%
        {toros2018homeless}
\bibfield{author}{\bibinfo{person}{Halil Toros} {and} \bibinfo{person}{Daniel
  Flaming}.} \bibinfo{year}{2018}\natexlab{}.
\newblock \showarticletitle{Prioritizing Homeless Assistance Using Predictive
  Algorithms: An Evidence-Based Approach}.
\newblock \bibinfo{journal}{\emph{Cityscape}} \bibinfo{volume}{20},
  \bibinfo{number}{1} (\bibinfo{year}{2018}).
\newblock
\urldef\tempurl%
\url{https://ssrn.com/abstract=3202479}
\showURL{%
\tempurl}


\bibitem[\protect\citeauthoryear{Turnell and Edwards}{Turnell and
  Edwards}{1997}]%
        {turnell1997aspiring}
\bibfield{author}{\bibinfo{person}{Andrew Turnell} {and} \bibinfo{person}{Steve
  Edwards}.} \bibinfo{year}{1997}\natexlab{}.
\newblock \showarticletitle{Aspiring to Partnership. The Signs of Safety
  approach to child protection}.
\newblock \bibinfo{journal}{\emph{Child Abuse Review: Journal of the British
  Association for the Study and Prevention of Child Abuse and Neglect}}
  \bibinfo{volume}{6}, \bibinfo{number}{3} (\bibinfo{year}{1997}),
  \bibinfo{pages}{179--190}.
\newblock


\bibitem[\protect\citeauthoryear{upEnd Movement}{upEnd Movement}{[n.d.]}]%
        {upEND}
\bibfield{author}{\bibinfo{person}{upEnd Movement}.}
  \bibinfo{year}{[n.d.]}\natexlab{}.
\newblock \bibinfo{title}{upEnd}.
\newblock \bibinfo{howpublished}{\url{https://upendmovement.org/}}.
\newblock
\newblock
\shownote{Last Accessed: May 10, 2022.}


\bibitem[\protect\citeauthoryear{upEND Movement}{upEND Movement}{2021}]%
        {endup2021}
\bibfield{author}{\bibinfo{person}{upEND Movement}.}
  \bibinfo{year}{2021}\natexlab{}.
\newblock \bibinfo{title}{How We endUP: A Future Without Family Policing}.
\newblock
\newblock
\urldef\tempurl%
\url{https://upendmovement.org/event/how-we-endup-a-future-without-family-policing/}
\showURL{%
\tempurl}
\newblock
\shownote{The second annual convening of the upEnd Movement.}


\bibitem[\protect\citeauthoryear{Vaithianathan, Dinh, Kalisher, Kithulgoda,
  Kulick, Mayur, Ning, and Benavides~Prado}{Vaithianathan
  et~al\mbox{.}}{2019}]%
        {douglascounty}
\bibfield{author}{\bibinfo{person}{Rhema Vaithianathan}, \bibinfo{person}{Haley
  Dinh}, \bibinfo{person}{Allon Kalisher}, \bibinfo{person}{Chamari
  Kithulgoda}, \bibinfo{person}{Emily Kulick}, \bibinfo{person}{Megh Mayur},
  \bibinfo{person}{Athena Ning}, {and} \bibinfo{person}{Diana
  Benavides~Prado}.} \bibinfo{year}{2019}\natexlab{}.
\newblock \bibinfo{title}{Implementing a Child Welfare Decision Aide in Douglas
  County | Methodology Report}.
\newblock
  \bibinfo{howpublished}{\url{https://csda.aut.ac.nz/__data/assets/pdf_file/0009/347715/Douglas-County-Methodology_Final_3_02_2020.pdf}}.
\newblock
\newblock
\shownote{Online; accessed 8-September-2021.}


\bibitem[\protect\citeauthoryear{Vaithianathan, Jiang, Maloney, Nand, and
  Putnam-Hornstein}{Vaithianathan et~al\mbox{.}}{2017}]%
        {vaithianathan2017}
\bibfield{author}{\bibinfo{person}{Rhema Vaithianathan}, \bibinfo{person}{Nan
  Jiang}, \bibinfo{person}{Tim Maloney}, \bibinfo{person}{Parma Nand}, {and}
  \bibinfo{person}{Emily Putnam-Hornstein}.} \bibinfo{year}{2017}\natexlab{}.
\newblock \bibinfo{title}{Developing Predictive Risk Models to Support Child
  Maltreatment Hotline Screening Decisions}.
\newblock
\newblock
\urldef\tempurl%
\url{https://www.alleghenycountyanalytics.us/wp-content/uploads/2017/04/Developing-Predictive-Risk-Models-package-with-cover-1-to-post-1.pdf}
\showURL{%
\tempurl}


\bibitem[\protect\citeauthoryear{Varshney, Park, Raji, Hiranandani,
  Harikrishna, Koyejo, Richardson, and Lee}{Varshney et~al\mbox{.}}{2021}]%
        {varshney2021participatory}
\bibfield{author}{\bibinfo{person}{Kush Varshney}, \bibinfo{person}{Tina Park},
  \bibinfo{person}{Inioluwa~Deborah Raji}, \bibinfo{person}{Gaurush
  Hiranandani}, \bibinfo{person}{Narasimhan Harikrishna},
  \bibinfo{person}{Oluwasanmi Koyejo}, \bibinfo{person}{Brianna Richardson},
  {and} \bibinfo{person}{Min~Kyung Lee}.} \bibinfo{year}{2021}\natexlab{}.
\newblock \bibinfo{title}{Participatory specification of trustworthy machine
  learning}.
\newblock
  \bibinfo{howpublished}{\url{https://www.abstractsonline.com/pp8/\#!/10390/session/446}}.
\newblock
\newblock
\shownote{Last Accessed: May 10, 2022.}


\bibitem[\protect\citeauthoryear{Waldfogel}{Waldfogel}{2009}]%
        {waldfogel2009prevention}
\bibfield{author}{\bibinfo{person}{Jane Waldfogel}.}
  \bibinfo{year}{2009}\natexlab{}.
\newblock \showarticletitle{Prevention and the child protection system}.
\newblock \bibinfo{journal}{\emph{The Future of Children}}
  \bibinfo{volume}{19}, \bibinfo{number}{2} (\bibinfo{year}{2009}),
  \bibinfo{pages}{195–--210}.
\newblock


\bibitem[\protect\citeauthoryear{Weiner and Chor}{Weiner and Chor}{2019}]%
        {dana2019predictive}
\bibfield{author}{\bibinfo{person}{Dana Weiner} {and} \bibinfo{person}{Brian
  Chor}.} \bibinfo{year}{2019}\natexlab{}.
\newblock \bibinfo{title}{Using Predictive Analytics to Inform Child Welfare
  Preventive Services in New York City, Predictive Analytics Forum 2019
  Presentation}.
\newblock
\newblock
\urldef\tempurl%
\url{https://predictive.cfrc.illinois.edu/pdf/NYC.pdf}
\showURL{%
\tempurl}


\bibitem[\protect\citeauthoryear{Whittaker, Crawford, Dobbe, Fried, Kaziunas,
  Mathur, West, Richardson, Schultz, and Schwartz}{Whittaker
  et~al\mbox{.}}{2018}]%
        {whittaker2018ai}
\bibfield{author}{\bibinfo{person}{Meredith Whittaker}, \bibinfo{person}{Kate
  Crawford}, \bibinfo{person}{Roel Dobbe}, \bibinfo{person}{Genevieve Fried},
  \bibinfo{person}{Elizabeth Kaziunas}, \bibinfo{person}{Varoon Mathur},
  \bibinfo{person}{Sarah~Mysers West}, \bibinfo{person}{Rashida Richardson},
  \bibinfo{person}{Jason Schultz}, {and} \bibinfo{person}{Oscar Schwartz}.}
  \bibinfo{year}{2018}\natexlab{}.
\newblock \bibinfo{booktitle}{\emph{AI now report 2018}}.
\newblock \bibinfo{publisher}{AI Now Institute at New York University}.
\newblock


\bibitem[\protect\citeauthoryear{Wolf, Zhu, Bullard, Lee, and Brubaker}{Wolf
  et~al\mbox{.}}{2018}]%
        {wolf2018changing}
\bibfield{author}{\bibinfo{person}{Christine~T. Wolf}, \bibinfo{person}{Haiyi
  Zhu}, \bibinfo{person}{Julia Bullard}, \bibinfo{person}{Min~Kyung Lee}, {and}
  \bibinfo{person}{Jed~R. Brubaker}.} \bibinfo{year}{2018}\natexlab{}.
\newblock \showarticletitle{The Changing Contours of "Participation" in
  Data-Driven, Algorithmic Ecosystems: Challenges, Tactics, and an Agenda}. In
  \bibinfo{booktitle}{\emph{Companion of the 2018 ACM Conference on Computer
  Supported Cooperative Work and Social Computing}} (Jersey City, NJ, USA)
  \emph{(\bibinfo{series}{CSCW '18})}. \bibinfo{publisher}{Association for
  Computing Machinery}, \bibinfo{address}{New York, NY, USA},
  \bibinfo{pages}{377–384}.
\newblock
\urldef\tempurl%
\url{https://doi.org/10.1145/3272973.3273005}
\showURL{%
\tempurl}


\bibitem[\protect\citeauthoryear{Wong}{Wong}{2020}]%
        {wong2020democratizing}
\bibfield{author}{\bibinfo{person}{Pak-Hang Wong}.}
  \bibinfo{year}{2020}\natexlab{}.
\newblock \showarticletitle{Democratizing Algorithmic Fairness}.
\newblock \bibinfo{journal}{\emph{Philosophy and Technology}}
  \bibinfo{volume}{33}, \bibinfo{number}{2} (\bibinfo{year}{2020}),
  \bibinfo{pages}{225--244}.
\newblock
\urldef\tempurl%
\url{https://doi.org/10.1007/s13347-019-00355-w}
\showDOI{\tempurl}


\bibitem[\protect\citeauthoryear{Zhu, Yu, Halfaker, and Terveen}{Zhu
  et~al\mbox{.}}{2018}]%
        {zhu2018value}
\bibfield{author}{\bibinfo{person}{Haiyi Zhu}, \bibinfo{person}{Bowen Yu},
  \bibinfo{person}{Aaron Halfaker}, {and} \bibinfo{person}{Loren Terveen}.}
  \bibinfo{year}{2018}\natexlab{}.
\newblock \showarticletitle{Value-sensitive algorithm design: Method, case
  study, and lessons}.
\newblock \bibinfo{journal}{\emph{Proceedings of the ACM on Human-Computer
  Interaction}} \bibinfo{volume}{2}, \bibinfo{number}{CSCW}
  (\bibinfo{year}{2018}), \bibinfo{pages}{1--23}.
\newblock


\end{thebibliography}

\newpage 

\appendix

\section{Participant Experiences \& Demographics}
\label{sec:demographics}

In this section, we describe the questions we asked participants about their demographics, participants' responses, and some justification for why we asked participants minimal questions about their child welfare involvement in the demographics section of the survey.

\subsection{Demographics Responses by Participant}
\label{sec:demographics-responses}

See Table~\ref{tab:participant-demographics} for all demographics.

\lsdelete{
\begin{table*}[h]
    \begin{tabular}{|p{1cm}|p{4cm}|p{4cm}|p{1.5cm}|p{1.5cm}|}
    \hline
        \textbf{ID} & \centering{\textbf{Personal experience}} & \centering{\textbf{Job experience}} & \textbf{Ever been investigated?} & \textbf{Social work experience / education?}\\ \hline
        \textbf{P1} & Former foster youth & ED of private CPS agency & - & Yes \\ \hline
        \textbf{P2} & - & PhD student & - & No \\ \hline
        \textbf{P3} & - & Attorney (family law \& ICWA) & - & No \\\hline
        \textbf{P4} & - & DHS licensor & Yes & Yes \\\hline
        \textbf{P5} & - & CPS worker & - & Yes \\\hline
        \textbf{P6} & - & PhD student & - & No \\\hline
        \textbf{P7} & - & CPS management & - & Yes \\\hline
        \textbf{P8} & - & Teacher (mandated reporter) & - & No \\\hline
        \textbf{P9} & - & CPS worker, lecturer & - & Yes \\\hline
        \textbf{P10} & - & Attorney & - & No \\\hline
        \textbf{P11} & - & Psychologist, attorney & - & No \\\hline
        \textbf{P12} & Impacted parent & - & Yes & No \\\hline
        \textbf{P13} & - & ED of private services agency & - & Yes \\\hline
        \textbf{P14} & - & CPS administrator & Yes & Yes \\\hline
        \textbf{P15} & - & CPS worker & No & Yes \\ \hline
        \textbf{P16} & - & CPS administrator & No & Yes \\ \hline
        \textbf{P17} & Transracial adoptee & MFT therapist & No & No \\\hline
        \textbf{P19} & - & CPS worker & No & Yes \\ \hline
        \textbf{P20} & Impacted parent & ED of parent advocacy group & Yes & No \\ \hline
        \textbf{P21} & Impacted parent & Assistant editor of parent education publication & Yes & No\\\hline
        \textbf{P22} & Impacted parent & parent trainer & Yes & No\\\hline
        \textbf{P23} & Impacted parent & parent advocate & Yes & Yes\\\hline
        \textbf{P24} & Impacted parent & parent advocate & Yes & Yes\\\hline
        \textbf{P25} & Impacted parent & parent advocate & Yes & Yes\\\hline
        \textbf{P26} & Impacted parent & - & - & - \\\hline
        \textbf{P27} & Impacted parent & Parent advocate & Yes & Yes\\\hline
        \textbf{P28} & Impacted parent & Parent advocate trainer & Yes & Yes\\\hline
        \textbf{P29} & - & Perinatal social worker & No & Yes\\\hline
        \textbf{P30} & - & CPS field director & Yes & Yes\\\hline
        \textbf{P31} & - & CPS project manager & No & Yes\\\hline
        \textbf{P32} & - & CPS worker & Yes & Yes\\\hline
        \textbf{P33} & Impacted parent & - & - \\\hline
        \textbf{P34} & Impacted parent & Parent advocate & Yes & Yes\\\hline
        \textbf{P35} & Impacted parent & Parent advocate & Yes & No\\\hline
        \textbf{P36} & Impacted parent & Parent advocate & - & -\\\hline
    \end{tabular}
    \caption{Participants' personal or occupational experiences with child welfare.}
    \label{tab:participant-experience}
\end{table*}
} % end lsdelete

\begin{table*}[h]
    \resizebox{17cm}{!}{%
    \begin{tabular}{|p{0.5cm}|p{7cm}|p{2cm}|p{1cm}|p{1.5cm}|p{2.5cm}|}
    \hline
        \textbf{ID} & \centering{\textbf{Race \& Ethnicity}} & \centering{\textbf{Gender \& Expression}} & \textbf{Age} & \textbf{Living Area} & \textbf{Highest Degree?}\\ \hline
        \textbf{P1} & white & Woman & 50-59 & Urban & MA or similar\\ \hline
        \textbf{P2} & white & Ciswoman & 25-29 & Suburban & Doctorate\\ \hline
        \textbf{P3} & Indigenous / Native American, white & Woman & 40-49 & Suburban & MA or similar\\ \hline
        \textbf{P4} & white & Woman & 50-59 & Suburban & Bachelor's\\ \hline
        \textbf{P5} & white & Woman & 30-39 & Suburban & MA or similar\\ \hline
        \textbf{P6} & white & Cisman & 18-24 & Urban & Doctorate\\ \hline
        \textbf{P7} & Black / African American, Hispanic / Latina / Latino / Latinx, Indigenous / Native American / Alaska Native / Native Hawaiian, mixed / multiracial & Woman & 40-49 & Urban & MA or similar\\ \hline
        \textbf{P8} & Asian & Woman & 25-29 & Urban & MA or similar\\ \hline
        \textbf{P9} & Black / African American, Hispanic / Latina / Latino / Latinx, white, mixed / multiracial & Woman & 30-39 & Urban, Suburban & MA or similar\\ \hline
        \textbf{P10} & white & Ciswoman & 50-59 & Urban & MA or similar\\ \hline
        \textbf{P11} & white & Man & 60+ & Suburban & MA or similar\\ \hline
        \textbf{P12} & Black / African American & Woman & 40-49 & Urban & MA or similar\\ \hline
        \textbf{P13} & white & Woman & 50-59 & Urban & MA or similar\\ \hline
        \textbf{P14} & Black / African American, mixed / multiracial & Woman & 50-59 & Urban & Doctorate\\ \hline
        \textbf{P15} & Black / African American & Woman & NA & Suburban & MA or similar\\ \hline
        \textbf{P16} & Black / African American & Ciswoman & 60+ & Suburban & MA or similar\\ \hline
        \textbf{P17} & Latinx, South American Native, mixed / multiracial & Non-binary, Genderqueer & 40-49 & Urban & MA or similar\\ \hline
        \textbf{P19} & Black / African American & Woman & 30-39 & Urban & Doctorate\\ \hline
        \textbf{P20} & Black / African American, Hispanic / Latina / Latino / Latinx & Woman & 40-49 & Urban & Associate's\\ \hline
        \textbf{P21} & Black / African American & Woman & 40-49 & NA & Associate's\\ \hline
        \textbf{P22} & Black / African American & Woman & 30-39 & Urban & Some, no degree\\ \hline
        \textbf{P23} & Black / African American & Woman & 50-59 & NA & Some, no degree\\ \hline
        \textbf{P24} & Black / African American & Woman & 40-49 & City & Some, no degree\\ \hline
        \textbf{P25} & mixed / multiracial & Woman & 40-49 & Urban & Associate's\\ \hline
        \textbf{P27} & Black / African American & Woman & 30-39 & City & Bachelor's\\ \hline
        \textbf{P28} & Hispanic / Latina / Latino / Latinx & Woman & 40-49 & Urban & Some, no degree\\ \hline
        \textbf{P29} & Black / African American, Indigenous / Native American / Alaska Native / Native Hawaiian & Woman & 30-39 & Suburban & Doctorate\\ \hline
        \textbf{P30} & Hispanic / Latina / Latino / Latinx & Woman & 30-39 & Rural & MA or similar\\ \hline
        \textbf{P31} & Black / African American & Woman & 40-49 & Suburban & Doctorate\\ \hline
        \textbf{P32} & Black / African American & Woman & 50-59 & Urban & MA or similar\\ \hline
        \textbf{P34} & Black / African American & Woman & 60+ & Urban & Some, no degree\\ \hline
        \textbf{P35} & Black / African American & Man & 40-49 & Suburban & Some, no degree\\ \hline
    \end{tabular}
    } % end resize
    \caption{Participants' self-disclosed demographics. When asked about race, ethnicity, gender identity, gender expression, participants were asked to choose as many or few options as they identified with. See Appendix~\ref{sec:demographics-questions} for exact survey questions and responses).}
    \label{tab:participant-demographics}
\end{table*}

\subsection{Demographics Questions}
\label{sec:demographics-questions}

We asked participants about their demographics during the pre-survey only. Below we include questions and answer options we asked participants:

\begin{adjustwidth}{2em}{0pt}

\noindent \textbf{How familiar are you with the Child Welfare system?}\\
\noindent Answer Options: unfamiliar, moderately unfamiliar, neither familiar nor unfamiliar, moderately familiar, familiar\\

\noindent \textbf{Do you work for a Child Welfare department or do you have an education in child social work?}\\
\noindent Answer Options: yes, no, prefer not to disclose\\

\noindent \textbf{What is your current occupation, if any? (If none, leave blank.)}\\
\noindent Answer Options: open response\\

\noindent \textbf{If you live in the U.S., which state do you live in? If you don't live in the U.S., which country do you live in?}\\
\noindent Answer Options: open response\\

\noindent \textbf{Do you live in a rural, urban, or suburban area?}\\
\noindent Answer Options: rural, urban, suburban, prefer not to disclose\\
	 	 	 		
\noindent \textbf{What is your age?}\\
\noindent Answer Options: 18 - 24, 25 - 34, 35 - 44, 45 - 54, 55 - 64, 65 - 74, 75 - 84, 85 or older, prefer not to disclose	\\

\noindent \textbf{What is the highest degree or level of schooling you have completed? If currently enrolled, the highest degree you are pursuing.}\\
\noindent Answer Options: no high school, some high school, high school diploma, some college, no degree, Associate’s degree, Bachelor’s degree, Master’s degree or professional program, Doctorate, prefer not to disclose\\

\noindent \textbf{Which of the following races/ethnicities do you identify as (please select all that apply)?}\\
Answer Options: Asian, Black / African American, Hispanic / Latina / Latino / Latinx; Indigenous / Native American / Alaska Native; white; mixed / multiracial; prefer to self-describe (+ open response), prefer not to disclose\\

\noindent \textbf{Which of the following gender identities/expressions do you identify as (please select all that apply)?}\\
\noindent Answer Options: woman, man, non-binary, transgender (current gender is different from what was assigned at birth), cisgender (current gender matches what was assigned at birth), prefer to self-describe (+ open response), prefer not to disclose\\

\noindent \textbf{Please let us know if you want us to know any other demographic information or experiences with the child welfare system that we didn't ask about.}\\
\noindent Answer Options: open response

\end{adjustwidth}

\subsection{Reasoning for voluntary disclosure of personal child welfare experiences}
\label{sec:voluntary-disclosure}
\noindent In order to better understand participants' personal experiences with the child welfare system, we: 1) asked participants how familiar with the system they were, 2) asked whether or not they have been subject to a child welfare investigation, and 3) we provided an open response question at the end of the survey and allowed participants to speak about their own experiences during the workshops if they so chose. If participants said they were unfamiliar with the system (which none did), their responses would not have been included in the study. We asked about participants' experiences in this way, rather than asking questions about more intrusive child welfare interactions, like \textit{``Have you ever had your children placed in foster care?''}, because we worried that explicitly asking about more intrusive interactions may have pressured some participants to describe or relive difficult or traumatizing experiences. This aligns with prior work on potential harms of personal disclosure in participatory workshops \cite{harrington2019deconstructing}. We also wanted to allow participants to define which experiences they thought were most relevant to this study within their own terms. There are tradeoffs and limitations to these approaches, however: Because we did not explicitly ask questions about the plethora of ways someone may be impacted by the child welfare system, there may be relevant experiences that additional participants had which they did not disclose to us. Two examples of questions that we did not ask about, but which reflect particularly relevant experiences, include those related to whether participants had experiences of themselves being in foster care or being adopted as youth.

\newpage

\section{Survey Questions \& Responses}
\label{sec:survey}
All but two questions in our pre-survey and post-survey were phrased as a statement which participants responded to with one of six options that they felt best represented their level of agreement with the statement: ``I strongly disagree''; ``I disagree''; ``I neither agree or disagree''; ``I agree''; ``I strongly agree''; or ``I prefer not to respond.'' Each of these questions was then followed up with an optional open response text that asked participants to explain why they answer that way to the immediately previous multiple choice question. The last two questions in the post-survey were optional open response questions not associated with a multiple choice question. To participants, we referred to the workshops as ``focus groups''and to PRMs as ``data-driven tools.'' The following questions were the first four questions in both the pre-survey and the post-survey:
\begin{enumerate}
    \item \textit{``I trust the current Child Welfare system to make decisions to prevent child maltreatment.''}
    \item \textit{``I trust the current Child Welfare system to make unbiased decisions.''}
    \item \textit{``I think using data-driven predictive tools to assist decision-making in Child Welfare will lead to better outcomes..''}
    \item \textit{``I think using data-driven predictive tools can help Child Welfare make more equitable decisions.''}
\end{enumerate}

The following questions were the last four questions in \textit{only} the post-survey (not in the pre-survey):
\begin{enumerate}
    \setcounter{enumi}{4}
     \item \textit{``I feel like the focus group helped me gain a better understanding of data-driven tools in Child Welfare''}
     \item \textit{``I feel like the focus group changed my views on using data-driven tools in Child Welfare''}
      \item \textit{``How did you like the focus group session?''}
      \item \textit{``What problems, if any, do you see with this process of group discussion about the Child Welfare system and the design and use of data-driven tools? (Optional)''}
\end{enumerate}

Almost no participants' responses to the first four questions above changed from the pre-survey (before the workshop) to the post-survey (after the workshop). Specifically, all participants' responses on questions 1, 2, and 4 were not changed and only 19 participants' responses on question 3 were changed from {``Neutral''} to either {``Strongly disagree''} or {``Disagree''}. Most participants responded to the third-to-last question of the post-survey saying that the workshop barely changed their views on PRMs in child welfare.

Although our results indicate that the focus-group workshop did not dramatically change participants' views on the current Child Welfare system or PRMs, participants generally said they enjoyed the workshop, and some said they learned from it. Specifically, 26 participants said they either ``Really liked'' or ``Liked'' the workshop. For example, P17 said that they liked the workshop because they were able to \textit{``hear from a wide range of professionals and learn from one another even though... everyone had some agreements and disagreements.''} P30 said that the workshop \textit{``helped [them] understand where the issue may start or areas that need more attention''}.
%However, participants generally said they enjoyed the workshop, and some said they learned from it. For example, P17 said that they liked the workshop because they were able to \textit{``hear from a wide range of professionals and learn from one another even though... everyone had some agreements and disagreements.''} P30 said that the workshop \textit{``helped [them] understand where the issue may start or areas that need more attention''}.

\section{Recruitment materials}
Below is a copy of the language in the call for participation we emailed out to recruit participants:

\begin{adjustwidth}{2em}{0pt}

\textbf{Study on Predictive Algorithms used in the Child Welfare system}

\noindent We are looking for stakeholders of the Child Welfare system, e.g. parents, community advocates, former youth who were in the system. You must be 18 or older to participate.

\noindent \textbf{Study:} We will hold 4-5 person focus groups to identify impacted stakeholders’ concerns with the use and design of predictive technologies in Child Welfare.

\noindent \textbf{Who we are:} We are students and researchers at Carnegie Mellon University, UC Berkeley, and University of Minnesota. We study societal impacts of new technologies.

\noindent \textbf{Background:} Child Welfare departments are starting to design and use technologies which use historical case data and possibly other county data to flag which kids are at higher risk. For example, Allegheny County uses the Allegheny Family Screening Tool.

\noindent \textbf{Details:} The study consists of two 5-10 minute surveys and a 90 minute focus group with a few other people. Compensation is \$50. Responses from the survey and focus group will be confidential, but may be anonymously included in a research paper and in Child Welfare policy suggestions.

\noindent If you would like to participate in this study, please fill out this interest form (\url{forms.gle/A8ac7kJYSSP9NExg8}) or email Logan Stapleton at lstaplet@andrew.cmu.edu with the heading \textit{Child Welfare Study}. Thank you!

\end{adjustwidth}

\lsdelete{
\xhdr{Solidarity with Impacted Families To Circumvent or oppose the Child Welfare System.}
In Section~\ref{sec:participant-concerns}, participants suggested researchers and designers work in solidarity with impacted families and communities in opposition to child welfare. Most prior work on child welfare in ML and HCI has been in partnership with CPS agencies or designed tools to help agencies \cite{chouldechova2018case,saxena2020participatory,saxena2021framework,brown2019toward,cheng2021soliciting,cheng2022disparities}.\lsdelete{Most of the authors of this paper have and continue to conduct research with CPS agencies.} Even in prior participatory work with impacted parents, \citet{brown2019toward} ask ``What can researchers and designers working \textit{in partnership} with public service agencies?'' and suggest us to ``[f]acilitate... positive relationships between child welfare workers and families.'' This might benefit families if agencies are actually helping them. Yet, many of our participants described an adversarial relationship between CPS agencies and impacted communities, especially poor, Black, and other marginalized communities. If researchers and designers only work to encourage positive relationships between families and agencies, we may alienate impacted communities who have been harmed and do not want to stay positive. Our paper is one of the first in ML or HCI to work with impacted families and communities independently of a child welfare agency.

Our participants suggest researchers and designers to continue this practice in the future.\lsdelete{\footnote{This aligns with the origins of participatory design work, which worked in solidarity marginalized communities \cite{bjorgvinsson2010participatory,spinuzzi2002scandinavian}}}

}

\lsdelete{Move to discussion: However, we did see some similarities in our findings: For one, \citet{brown2019toward} find that their participants had low comfort with PRMs because of ``system-level concerns.'' In a similar vein, our participants were worried that PRMs would perpetuate or exacerbate existing problems in the child welfare system. For another, both our sets of participants were very concerned about ``bias on the part of case workers involved in the decision process, as well as bias present in the data or the algorithm.'' However, }

\lsdelete{If one believes that agencies are truly helping families, working to help them may be done with good intentions. It is also often easiest for researchers to partner with agencies: they are the largest, most funded, most organized entities in the child welfare system. However, this stature creates a power imbalance between agencies and communities. As our results indicate, many families and communities often feel distrust, suspicion, fear, or direct opposition towards agencies. ML and HCI researchers' work in partnership with agencies may alienate many impacted families and communities.  It may seed doubt and suspicion of the researchers themselves.  This study is one of the first studies in ML or HCI to interact with impacted families and communities independently of a child welfare agency. Our participants offer not only a new methodological approach, but specific suggestions for researchers and designers. }

\lsdelete{Recommendations in Section~\ref{sec:results} should be both as design ideas in themselves, and reflection on impacted stakeholders' perceptions of the child welfare system. For instance, the suggestion to abolish the child welfare system can be taken both as a literal call to tear it down, and as a reflection that many of the impacted stakeholders that we spoke with feel that the child welfare system is so harmful that they would rather see it gone.}

\end{document}